\documentclass[10pt,journal,final,twocolumn]{IEEEtran}
\usepackage{epsfig,graphicx,subfigure,psfrag,amsmath,cases,bm}
\usepackage{latexsym,amssymb,algorithm,mathtools}
\usepackage{algorithmic}
\usepackage{color}
\usepackage{url}
\usepackage{scrtime}
\usepackage{stfloats}
\usepackage{tablefootnote}
\usepackage{cite}
\usepackage{amsfonts,mathabx}
\usepackage{textcomp}
\usepackage{xcolor,capt-of}
\usepackage{multirow}
\usepackage{amsthm}
\usepackage{subfigure}
\usepackage{makecell}
\usepackage{pifont}
\usepackage{amssymb}
\usepackage[utf8]{inputenc}
\usepackage[T1]{fontenc}
\usepackage{geometry}
\allowdisplaybreaks

\newtheorem{remark}{Remark}
\newtheorem{lemma}{Lemma}

\newtheorem{T-Prob}{Transformed Problem}
\newtheorem{Prop}{Proposition}

\DeclareMathOperator{\maxo}{maximize}
\DeclareMathOperator{\mino}{minimize}
\DeclareMathOperator{\diag}{\mathrm{diag}}

\DeclareMathOperator{\subto}{subject\hspace*{2mm}to}

\newcommand{\QED}{\hfill \ensuremath{\blacksquare}}

\newcommand{\myoverline}[1]{\overline{\overline{#1}}}
\newcommand{\mywidetilde}[1]{\widetilde{\widetilde{#1}}}

\begin{document} 

\title{Joint Radiation Power, Antenna Position, and Beamforming Optimization for Pinching-Antenna Systems with Motion Power Consumption}

\author{
    Yiming Xu, \textit{Graduate Student Member, IEEE,}  Dongfang Xu, \textit{Member, IEEE,} Xianghao Yu, \textit{Senior Member, IEEE,} \\ Shenghui Song, \textit{Senior Member, IEEE,} Zhiguo Ding, \textit{Fellow, IEEE,} and  Robert Schober, \textit{Fellow, IEEE}
\vspace{-9mm}
\thanks{Yiming Xu, Dongfang Xu, and Shenghui Song are with the Hong Kong University of Science and Technology, Hong Kong, China (email: yxuds@connect.ust.hk, \{eedxu, eeshsong\}@ust.hk); 
Xianghao Yu is with the City University of Hong Kong, Hong Kong, China (email: alex.yu@cityu.edu.hk); Zhiguo Ding is with the University of Manchester, Manchester, England (email: zhiguo.ding@manchester.ac.uk); Robert Schober is with the Institute for Digital Communications, Friedrich-Alexander-University Erlangen-N{\"u}rnberg, Erlangen, Germany (email: robert.schober@fau.de).
}
}

\maketitle
\begin{abstract}
Pinching-antenna systems (PASS) have been recently proposed to improve the performance of wireless networks by reconfiguring both the large-scale and small-scale channel conditions. However, existing studies ignore the physical constraints of antenna placement and assume fixed antenna radiation power. To fill this research gap, this paper investigates the design of PASS taking into account the motion power consumption of pinching-antennas (PAs) and the impact of adjustable antenna radiation power.
To that end, we minimize the average power consumption for a given quality-of-service (QoS) requirement, by jointly optimizing the antenna positions, antenna radiation power ratios, and transmit beamforming. To the best of the authors' knowledge, this is the first work to consider radiation power optimization in PASS, which provides an additional 
degree of freedom (DoF) for system design. The cases with both continuous and discrete antenna placement are considered, where the main challenge lies in the fact that the antenna positions affect both the magnitude and phase of the channel coefficients of PASS, making system optimization very challenging.
To tackle the resulting unique obstacles, an alternating direction method of multipliers (ADMM)-based framework is proposed to solve the problem for continuous antenna movement, while its discrete counterpart is formulated as a mixed integer nonlinear programming (MINLP) problem and solved by the block coordinate descent (BCD) method.
Simulation results validate the performance enhancement achieved by incorporating PA movement power assumption and adjustable radiation power into PASS design, while also demonstrating the efficiency of the proposed optimization framework. The benefits of PASS over conventional multiple-input multiple-output (MIMO) systems in mitigating the large-scale path loss and inter-user interference is also revealed.
\end{abstract}

\begin{IEEEkeywords}
Beamforming, antenna placement and activation, motion power consumption, antenna radiation power, pinching-antenna system (PASS), alternating direction
method of multipliers (ADMM).
\end{IEEEkeywords}

\section{Introduction}
Over the past few decades, multiple-input multiple-output (MIMO) has evolved as an effective technique for enhancing the throughput, spectral efficiency, and reliability of wireless communication systems \cite{6798744}. However, in traditional MIMO systems, the antennas are deployed in fixed locations, limiting the channel capacity.
To break through this limitation, flexible-antenna techniques, such as reconfigurable intelligent surfaces (RISs) \cite{9690635}, movable antennas \cite{10906511}, and fluid antennas \cite{9264694}, have been developed. Specifically, an RIS consists of a large number of low-cost electromagnetic reflecting elements that can intelligently reconfigure the wireless channels. The movable and fluid antenna systems enhance the channel quality by optimizing the antenna positions and orientations.
Despite their proven effectiveness, existing flexible-antenna techniques face practical limitations when it comes to reducing large-scale path loss and maintaining line-of-sight (LoS) links, which are particularly critical for future millimeter-wave (mmWave) and Terahertz (THz) communications.
For instance, although an RIS is able to construct a virtual LoS link, RIS-aided systems suffer from high path loss due to the double-fading effect \cite{9690635}. On the other hand, movable and fluid antennas typically move within the range of a few wavelengths, and thus are unable to mitigate LoS blockage.
\par
To this end, pinching-antenna systems (PASS) have recently emerged as a revolutionary flexible-antenna technology promising to overcome the aforementioned obstacles \cite{10945421,liu2025pinching}. PASS was initially proposed and demonstrated by NTT DOCOMO in 2022 \cite{suzuki2022pinching}. Specifically, PASS employs dielectric waveguides as transmission medium, spanning from a few meters to tens of meters, along which multiple radiation points can be activated by applying small and cost-effective dielectric particles, such as cloth pinches used in DOCOMO's testbed \cite{suzuki2022pinching}. These dielectric particles are referred to as \textit{pinching antennas} (PAs), through which the signals in the waveguide are emitted into free space. PAs can be flexibly activated, deactivated, or repositioned along the waveguide, enabling the system to reconfigure the channels on both large and small scales. 
\par
Compared with conventional MIMO systems and other flexible-antenna technologies, PASS offer several unique advantages: \textbf{1) Path loss mitigation}: Since the PAs can be moved over a wide range on the waveguide, PASS can activate the PAs near the users to significantly reduce the path loss.
\textbf{2) LoS link establishment}: By dynamically adjusting the positions of PAs to bypass a blockage, PASS can construct and robustly maintain a LoS link.
\textbf{3) Scalable system deployment}: The low-cost PAs can be easily reconfigured by pinching or releasing operations, making the deployment of PASS very scalable.
\textbf{4) Efficient \textit{pinching beamforming}}: PASS facilitate a novel beamforming paradigm, \textit{pinching beamforming}, which simultaneously adjusts both the path loss and the phases experienced by the radiated signals. Benefiting from the pinching beamforming, PASS achieve efficient signal enhancement and interference mitigation, even with a limited number of PAs \cite{xu2025joint}.

\par

\par
Inspired by the aforementioned advantages, a significant amount of research has been conducted regarding the performance analysis and system design of PASS. In \cite{10945421}, the authors analyzed the ergodic sum rate achieved by PASS, revealing the superior gain of PAs over conventional antennas. Furthermore, a performance upper bound on the multi-input single-output (MISO) interference channel with two users was derived.
In \cite{10896748}, the authors proposed a two-stage algorithm to optimize the antenna positions. Specifically, the antenna positions were first optimized to minimize the path loss and then refined to maximize the receive signal strength. 
The authors of \cite{10912473} formulated the antenna position optimization problem for PASS with one waveguide as a one-sided one-to-one matching problem based on matching theory. 
\par
Besides antenna placement, the antenna radiation power also plays an important role in PASS. The first physical radiation model for PASS was proposed in \cite{wang2025modeling}, unveiling the relation between radiation power and antenna deployment. The authors of \cite{xu2025pinching} proposed a physical model to configure antenna radiation power ratios by tuning the spacing between PAs and waveguides. A closed-form expression for the spacing between PAs and waveguides was derived to achieve equal radiation power for all antennas. Based on that, a global optimal algorithm was developed for minimization of the transmit power of PASS with discrete PAs movement.
\par
Despite these recent advancements, there are still many challenges that need to be tackled to fully unleash the potential of PASS. First, the physical constraints of antenna movement were ignored by existing works when designing the PA positions. On the one hand,
the power consumption of PA movement, controlled mechanically or electronically, has not been considered in the literature. On the other hand, in practice, the movement distance of PAs within a given time is limited, but this constraint has not been fully considered in existing works.
\par
Another issue that has not been well investigated is antenna radiation power control. In particular, most of the existing works assumed equal antenna radiation power \cite{qin2025joint,zhu2025pinching,10912473}, hindering the PASS to reach its full potential. Although the authors of \cite{wang2025modeling} and \cite{xu2025pinching} have made initial contributions in modeling the antenna radiation power, the radiation power optimization in PASS has not been addressed in the literature, so far.
\par
To bridge the above research gaps, this paper investigates the design of PASS taking the physical movement constraints on PAs and the adjustable radiation power into consideration. Specifically, we consider a downlink multi-user MIMO PASS, where the antenna movement, antenna radiation power, and transmit beamforming are jointly optimized to minimize the average power consumption, while guaranteeing the users' quality-of-service (QoS) requirements. 
Furthermore, cases with both continuous and discrete antenna movement are investigated. 
\par
The challenge in solving the formulated optimization problem for continuous antenna movement stems from two aspects. First, the challenging spherical-wave channel model needs to be used in PASS, instead of the simplified far-field beam-steering model commonly used for conventional flexible-antenna system design. For example, while the antenna positions affect only the phase of the channel coefficients for movable and fluid antenna systems, the antenna positions in PASS also impact the magnitude of the channel coefficients, causing challenges for optimization. In addition, the optimization variables for transmit beamforming, antenna radiation power, and channel coefficients, which are complicated functions of the antenna positions, are severely coupled. To this end,
an optimization framework based on the alternating direction
method of multipliers (ADMM) is developed to solve the problem for continuous antenna movement. The problem for discrete antenna movement is treated as a PA position selection problem, resulting in a mixed integer nonlinear program (MINLP). 
The main difficulties in solving the problem for discrete antenna movement are caused by the non-convex binary constraint and variable coupling, where a block coordinate descent (BCD)-based framework is proposed to tackle the problem. 
\par
The main contributions of this paper can be summarized as follows.
\begin{itemize}
    \item We investigate a PASS accounting for the physical constraints on PA movement and adjustable radiation power. A novel comprehensive framework for joint antenna movement, antenna radiation power, and transmit beamforming design is proposed for minimization of the average power consumption, including both information transmission power and antenna movement power, while meeting the users' signal-to-interference-plus-noise ratio (SINR) requirements.
    \item To tackle the unique challenges arising in optimizing the PASS, an ADMM-based framework is developed to solve the problem for continuous antenna movement. Specifically, we adopt the variable splitting method to decompose the complex impact of the antenna positions on the channel coefficients, resulting in several equality constraints. Then, we adopt ADMM to handle these equality constraints and the variable coupling issue by alternatingly minimizing the augmented Lagrangian function. By exploiting a series of transformations, fractional programming, and successive convex approximation (SCA), the resulting sub-problems are effectively solved.
    \item
    To handle the discrete antenna movement case, we formulate the problem as a binary selection task for each PA among multiple discrete position choices, leading to an MINLP problem. Then, a BCD-based algorithm is developed to solve the problem. Specifically, bilinear transformation is exploited to decouple the binary variables and the transmit beamforming. Further, we use the penalty method and SCA to obtain a low-complexity design for handling the binary constraint.
    \item Numerical results are presented to validate the effectiveness of incorporating the physical constraints on PA movement and the adjustable radiation power into PASS design, and to evaluate the efficiency of the proposed optimization framework. Our results demonstrate that 1) it is essential to strike a balance between the powers consumed for information transmission and antenna movement, respectively; 2) adjusting the antenna radiation power ratios further improves the performance of PASS; 3) compared with conventional MIMO, PASS are superior in terms of both path loss reduction and interference management.
\end{itemize}





\par


\par
\textit{Notations:} 
Vectors and matrices are denoted by boldface lowercase and boldface capital letters, respectively. 
$(\cdot)^T$ and $(\cdot)^H$ stand for the transpose and the conjugate transpose operator, respectively. $\Re\{\cdot\}$ and $\Im\{\cdot\}$ represent the real and imaginary parts of a complex number, respectively. Vectorization of matrix $\mathbf{A}$ is denoted by $\mathrm{vec}(\mathbf{A})$, and $\mathbf{A}\otimes\mathbf{B}$ represents the Kronecker product between two matrices $\mathbf{A}$ and $\mathbf{B}$. $\angle (a)$ denotes the phase angle of the complex scalar $a$.

\begin{figure}[t]
\centering
\includegraphics[width=2.8in]{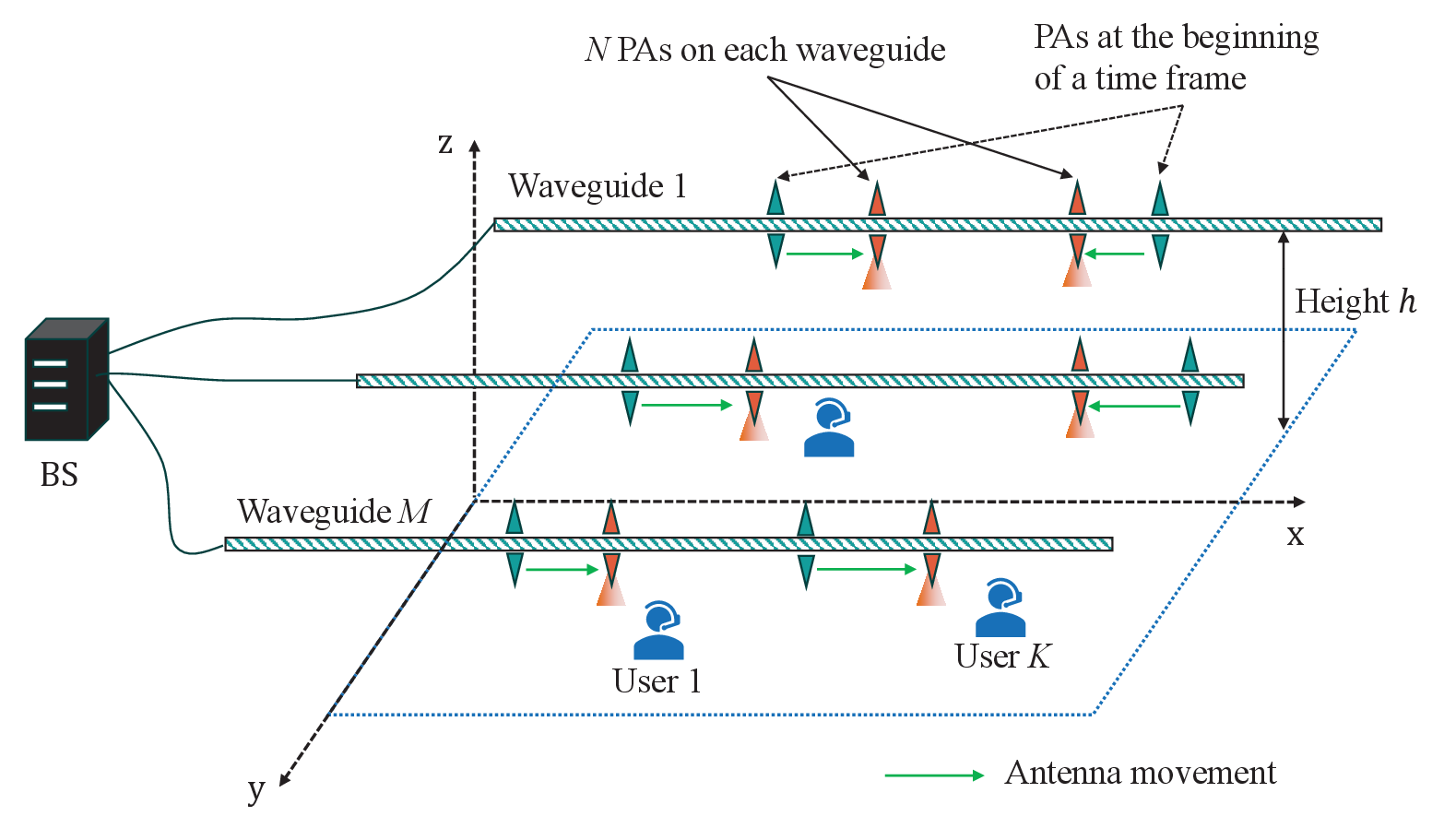}
\vspace*{-4mm}
\caption{Illustration of a PASS-enabled downlink multi-user MIMO communication system.}
\label{figure:illus}
\vspace*{-1mm}
\end{figure}




\section{System Model and Problem Formulation}
In this paper, we consider a PA-enabled downlink MIMO system, as shown in Fig. \ref{figure:illus}. 
The channel between PAs and users is dominated by a strong LoS link, which is dependent on the user's position \cite{wang2025modeling}. We adopt a two-phase protocol for configuring the PASS, where we assume that the user's position remains almost unchanged within each time frame \cite{wei2025mechanical}. In the first phase, the PAs move to the desired position identified by optimization. Then, the information transmission with optimized transmit beamforming is implemented in the second phase. We use $T_1$ and $T_2$ to denote the durations of the two phases, respectively. In an ideal case, the antennas can be precisely moved to any location, leading to a continuous antenna movement design. However, in practice, due to limited precision, the motion control of electromechanical drivers is confined to discrete movement \cite{8060521}. For a comprehensive investigation of antenna movement control, we consider both continuous and discrete antenna movement. As shown in Fig. \ref{figure:illus}, the PAs are preinstalled on waveguides and then moved to the designated position. For the discrete antenna movement case, the positions that the PAs can reach are quantized.
\par
The BS is equipped with $M$ waveguides, each of which is fed by an RF chain and comprises $N$ PAs. Without loss of generality, the waveguides are assumed to be deployed parallel to the $x$-axis at a height of $h$, as shown in Fig. \ref{figure:illus}. The $y$-axis coordinate of the $m$-th waveguide is denoted by $y_m$. The $m$-th waveguide spans from its feed point $\mathbf{p}_{m,0} = [0,y_m,h]^T$ to $[D,y_m,h]^T$ with a length of $D$. The coordinate of the $n$-th PA on the $m$-th waveguide is given by $\mathbf{p}_{m,n} = [x_{m,n}, y_m, h]^T$, where $x_{m,n}$ denotes the $x$-axis coordinate of the $n$-th PA on the $m$-th waveguide. We assume $x_{m,1} < x_{m,2} < \cdots < x_{m,N}$ without loss of generality. The coordinate of the $n$-th PA on the $m$-th waveguide at the beginning of a time frame is denoted by $\mathbf{p}'_{m,n} = [x'_{m,n}, y_m, h]^T$. The PASS serve $K$ ground users, and the coordinate of the $k$-th ground user is denoted by $\overline{\mathbf{p}}_k = [\overline{x}_k, \overline{y}_k, 0]^T$. For notational simplicity, we define sets $\mathcal{M}= \{ 1, \dots, M \}$, $\mathcal{N}= \{ 1, \dots, N \}$, and $\mathcal{K}= \{ 1, \dots, K \}$ to collect the indices of the waveguides, PAs, and users, respectively. Besides, we use the tuple $(m,n)$ to index the $n$-th PA on the $m$-th waveguide.
\subsection{Signal Model}
Denote the vector collecting the information symbols intended for the $K$ users by $\mathbf{s} = [s_1, \dots, s_K]^T \in \mathbb{C}^{K \times 1}$ with $\mathbb{E} [\mathbf{s} \mathbf{s}^H ] = \mathbf{I}_K$ and the transmit beamformer for the $k$-th user by $\mathbf{w}_k = [w_{1,k}, \dots, w_{M,k}]^T \in \mathbb{C}^{M \times 1}$. The feed signal at the $m$-th waveguide is given by
\begin{align} \label{signal_1}
c_m = \underset{k \in \mathcal{K}}{\sum} w_{m,k} s_k.
\end{align}
We assume lossless in-waveguide propagation because the in-waveguide attenuation has been shown to be insignificant \cite{10912473}.
After the in-waveguide propagation, the signal transmitted from the $N$ PAs on the $m$-th waveguide is given by $\mathbf{g}_m c_m$, where $\mathbf{g}_m \in \mathbb{C}^{N \times 1}$ denotes the in-waveguide channel vector given by
\begin{align}
\mathbf{g}_m \overset{\triangle}{=} [ \alpha_{m,1} e^{-j\frac{2 \pi}{\lambda_g}\|\mathbf{p}_{m,1} - \mathbf{p}_{m,0} \|}, \dots, \alpha_{m,N} e^{-j\frac{2 \pi}{\lambda_g}\|\mathbf{p}_{m,N} - \mathbf{p}_{m,0} \|}]^T.
\end{align}
Here, $\alpha_{m,n}$ denotes the radiation power ratio at the $(m,n)$-th PA, satisfying $\underset{n \in \mathcal{N}}{\sum} \alpha_{m,n}^2 \leq 1, \hspace{1mm} \forall m \in \mathcal{M}$, where, $\alpha_{m,n}=0$ indicates that the $(m,n)$-th antenna is deactivated.
Constant $\lambda_g = \frac{\lambda_c}{n_{\text{eff}}}$ represents the waveguide wavelength, with $\lambda_c$ denoting the free-space wavelength and $n_{\text{eff}}$ representing the effective refractive index of the waveguide \cite{pozar2012microwave}. 
\begin{remark}
The existing works in PASS design assume fixed radiation powers, which limits the PASS in fully exploiting its capabilities. In this work, we consider PASS design with adjustable radiation power, providing a new degree of freedom (DoF) for system optimization and allowing a more flexible power allocation.
\end{remark}
For the ease of presentation, $\mathbf{g}_m$ is reformulated as
\begin{equation}
    \mathbf{g}_m = \mathbf{G}_m \bm{\alpha}_m,
\end{equation}
where $\mathbf{G}_m \in \mathbb{C}^{N \times N}$ and $\bm{\alpha}_m \in \mathbb{R}^{N \times 1}$ are given by, respectively,
\begin{align}
\mathbf{G}_m &= \mathrm{diag}\big([ e^{-j\frac{2 \pi}{\lambda_g}\|\mathbf{p}_{m,1} - \mathbf{p}_{m,0} \|}, \dots, e^{-j\frac{2 \pi}{\lambda_g}\|\mathbf{p}_{m,N} - \mathbf{p}_{m,0} \|}]\big), \\
\bm{\alpha}_m & = [\alpha_{m,1}, \dots, \alpha_{m,N}]^T.
\end{align}
The received signal at the $k$-th user is given by
\begin{align} \label{received_signal}
y_k = \underset{m \in \mathcal{M}}{\sum} \mathbf{h}_{m,k}^H \mathbf{g}_m c_m + n_k,
\end{align}
where $\mathbf{h}_{m,k} \in \mathbb{C}^{N \times 1}$ is the free-space channel vector between the antennas of the $m$-th waveguide and the $k$-th user, and is given by
\begin{align}
    \mathbf{h}_{m,k} = \left[\frac{\beta e^{-j\frac{2\pi}{\lambda_c} \|\mathbf{p}_{m,1} - \overline{\mathbf{p}}_k \| }}{\|\mathbf{p}_{m,1} - \overline{\mathbf{p}}_k \|}, \dots, \frac{\beta e^{-j\frac{2\pi}{\lambda_c} \|\mathbf{p}_{m,N} - \overline{\mathbf{p}}_k \| }}{\|\mathbf{p}_{m,N} - \overline{\mathbf{p}}_k \|} \right]^H.
\end{align}
Here, $\beta = \frac{\lambda_c}{4 \pi}$ is a constant and $n_k \sim \mathcal{CN}(0,\sigma_k^2)$ denotes the additive white Gaussian noise (AWGN) at the $k$-th user. By defining 
\begin{align}
\mathbf{h}_k & \overset{\triangle}{=} \left[\mathbf{h}_{1,k}^H, \dots, \mathbf{h}_{M,k}^H \right]^H \in \mathbb{C}^{MN \times 1}, \\
\mathbf{G} & \overset{\triangle}{=} \diag \left[ \mathbf{G}_1, \dots, \mathbf{G}_M \right] \in \mathbb{C}^{MN \times MN}, \\
\mathbf{A} & \overset{\triangle}{=} \diag \left[ \bm{\alpha}_1, \dots, \bm{\alpha}_M \right] \in \mathbb{R}^{MN \times M}, \label{A_expression} \\
\mathbf{c} & \overset{\triangle}{=} [c_1, \dots, c_M]^T \in \mathbb{C}^{M \times 1},
\end{align}
the received signal $y_k$ can be rewritten compactly as
\begin{align}
y_k & = \mathbf{h}_k^H \mathbf{G} \mathbf{A} \mathbf{c} + n_k \notag \\
& = \underbrace{\mathbf{h}_k^H\mathbf{G} \mathbf{A} \mathbf{w}_k s_k}_{\text{Desired signal}} + \underbrace{\underset{k' \in \mathcal{K}\setminus \{k\}}{\sum} \mathbf{h}_k^H\mathbf{G} \mathbf{A} \mathbf{w}_{k'} s_{k'}}_{\text{Multi-user interference}} + n_k.
\end{align}
As a result, the SINR at the $k$-th user is given by
\begin{align}
\gamma_k = \frac{\left| \mathbf{h}_k^H\mathbf{G} \mathbf{A} \mathbf{w}_k \right|^2}{ \underset{k' \in \mathcal{K}\setminus \{k\}}{\sum}\left| \mathbf{h}_k^H\mathbf{G} \mathbf{A} \mathbf{w}_{k'} \right|^2 + \sigma_k^2}.
\end{align}
\subsection{Pinching-Antenna Movement Model}
The movement of the PAs is driven by electric motors, such as stepper motors \cite{8060521}. Denote the power consumption of the employed motor by $P$ and the speed of the PAs by $v$. Then, the maximum movement distance of one PA during each frame is given by $v T_1$, resulting in the following constraint on PA movement
\begin{align}
\left\| \mathbf{p}_{m,n} - \mathbf{p}'_{m,n} \right\| \leq v T_1.
\end{align}
As a result, the energy consumption of the $(m,n)$-th PA for moving from $\mathbf{p}'_{m,n}$ to $\mathbf{p}_{m,n}$ is obtained as follows \cite{wu2024globally}
\begin{align}
E_{m,n} = P \frac{\left\| \mathbf{p}_{m,n} - \mathbf{p}'_{m,n} \right\|}{v}.
\end{align}
\par
\begin{remark}
The existing literature on PASS mostly relies on antenna activation, which leads to suboptimal performance, since the PAs have to be preinstalled and their positions may not be optimal. A more general approach is to first activate the antenna closest to the optimal position and then move it to that position on a small scale. Such a PASS still adheres to the constraints on maximum movement distance and motion power consumption and is covered by the considered system model.
\end{remark}

\subsection{Problem Formulation for Continuous Antenna Movement}
In this paper, we aim to minimize the average power consumption while guaranteeing the communication QoS requirements of the users by jointly optimizing the radiation power ratios, antenna movement, and transmit beamforming. The resulting optimization problem for continuous antenna movement is formulated as follows
\begin{align} \label{continuous}
\underset{\substack{ \mathbf{w}_k, \bm{\alpha}_m, \mathbf{X} }}{\mino} \hspace*{3mm} & f_1 \overset{\triangle}{=} \frac{T_2}{T_1 + T_2} \underset{k \in \mathcal{K}}{\sum} \| \mathbf{w}_k\|^2_2 + \frac{1}{T_1 + T_2} \underset{\substack{m, n}}{\sum} E_{m,n} \notag\\
\subto\hspace*{2mm} & \mbox{C1:} \hspace{1mm} \gamma_k \geq \Gamma_k, \hspace{1mm} \forall k,\notag \\
& \mbox{C2:}\hspace*{1mm} \left\| \bm{\alpha}_m \right\|^2_2 \leq 1, \hspace{1mm} \forall m,\notag \\
& \mbox{C3:}\hspace*{1mm} x_{m,n} - x_{m,n-1} \geq \Delta_{\mathrm{min}}, \forall \hspace*{1mm} 2 \leq n \leq N, \forall m, \notag \\
& \mbox{C4:}\hspace*{1mm} 0 \leq x_{m,n} \leq D, \hspace*{1mm} \forall n, \hspace{1mm} \forall m, \notag \\
& \mbox{C5:}\hspace*{1mm} | x_{m,n} - x'_{m,n} | \leq v T_1, \hspace*{1mm} \forall n, \hspace{1mm} \forall m.
\end{align}
Here, we define matrix $\mathbf{X} \in \mathbb{R}^{M \times N}$ to collect the locations of all PAs, where its $(m, n)$-th element is $x_{m,n}$.
Constraint C1 ensures that the SINR of the $k$-th user is larger than the pre-defined threshold $\Gamma_k$. Constraint C2 indicates that the radiated power of each waveguide does not exceed the power of the feed signal. To avoid mutual coupling among antennas, we restrict the minimum distance between adjacent antennas to be larger than the pre-defined threshold $\Delta_{\mathrm{min}}$, as specified in constraint C3 \cite{10945421}. Constraint C4 ensures that the PAs move on the waveguide within the waveguide length. Constraint C5 guarantees that the movement distance of PAs is smaller than the maximum possible movement distance. We note that the optimization of the radiation power coefficients, i.e., $\bm{\alpha}_m$, has not yet been studied for PASS design.
\par
\begin{remark}
The optimization problem in \eqref{continuous} is highly non-convex. In particular, the transmit beamforming vector $\mathbf{w}_k$, antenna radiation power coefficient $\bm{\alpha}_m$, and antenna position matrix $\mathbf{X}$ are coupled. Different from movable antenna and fluid antenna systems, where the antenna positions only affect the phase of channel coefficients \cite{10354003,10753482}, the positions of PAs also impact the path loss, making the optimization problem more challenging.
\end{remark}

\section{Joint Power and Antenna Control for Continuous Antenna Movement}
In this section, an optimization framework based on ADMM is proposed to solve the formulated problem for continuous antenna movement. 


\begin{remark}
ADMM is well-suited for solving optimization problems with equality constraints and coupled variables, rendering it effective in dealing with the structure of the specific problem obtained for PASS design. 
Specifically, decomposing the channel coefficients, which are a complicated expression of the antenna positions, into multiple simpler expressions using the variable splitting method inherently introduces equality constraints, i.e., C6 and C7 specified below. Following the principle of augmented Lagrangian methods, ADMM can effectively handle these equality constraints by combining the Lagrangian term with quadratic penalty terms, as formulated in \eqref{lagrange_1} and elaborated on below \cite{boyd2011distributed}. In addition, ADMM partitions the initial problem into simpler sub-problems by splitting the variables and alternatingly solving these sub-problems to update the variables, enabling ADMM to overcome the variable coupling issue.
\end{remark}
\par
\subsection{ADMM-based Transformation}
In the following, we use the variable splitting method to transform the equivalent channel vector $\mathbf{G}^H \mathbf{h}_k$ of user $k$ into a more tractable form \cite{5427079}. Specifically, the $((m-1)N+n)$-th element of $\mathbf{G}^H \mathbf{h}_k \in \mathbb{C}^{MN \times 1}$ is given by
\begin{align}
[\mathbf{G}^H \mathbf{h}_k]_{((m-1)N+n)} = \frac{\beta e^{j\frac{2\pi}{\lambda_c}(r_{k,m,n}+n_{\text{eff}}x_{m,n})}}{r_{k,m,n}},
\label{channelexponent}
\end{align}
where $r_{k,m,n}$ denotes the distance between the $(m,n)$-th PA and the $k$-th user, which is given by
\begin{align}
r_{k,m,n} = \sqrt{(x_{m,n} - \overline{x}_k)^2+(y_m-\overline{y}_k)^2+h^2}.
\end{align}
To handle the exponential term in \eqref{channelexponent}, we define an auxiliary variable $\bm{\theta} \in \mathbb{R}^{KMN \times 1}$ and introduce the following equality constraint
\begin{align}
\mbox{C6:} \hspace{1mm} \theta_{k,m,n} = \frac{2 \pi}{\lambda_c} (r_{k,m,n}+n_{\text{eff}}x_{m,n}), \hspace*{1mm}\forall k, \hspace*{1mm}\forall m, \hspace*{1mm}\forall n,
\end{align}
where $\theta_{k,m,n}$ is the $((k-1)MN + (m-1)N + n)$-th element of $\bm{\theta}$.
For ease of presentation, we define vector $\bm{\phi} \in \mathbb{R}^{KMN \times 1}$, whose $((k-1)MN + (m-1)N + n)$-th element is $\phi_{k,m,n} \overset{\triangle}{=} \frac{2 \pi}{\lambda_c} (r_{k,m,n}+n_{\text{eff}}x_{m,n})$, and rewrite constraint C6 compactly as
\begin{align} \label{C6}
\mbox{C6:} \hspace{1mm} \bm{\theta} = \bm{\phi}.
\end{align}
Furthermore, we introduce auxiliary variables $\mathbf{t}_k \in \mathbb{C}^{MN \times 1}$ to replace the equivalent channel vector, leading to the following equality constraint
\begin{align}
\mbox{C7:} \hspace{1mm} \mathbf{t}_k = \mathbf{G}^H \mathbf{h}_k, \hspace{1mm}\forall k \in \mathcal{K}.\label{vectort_k}
\end{align}
The element-wise expression of constraint C7 is given by
\begin{align} \label{t_1}
t_{k,m,n} = \frac{\beta e^{j \theta_{k,m,n} }}{r_{k,m,n}}, \hspace{1mm} \forall k \in \mathcal{K}, \hspace{1mm}\forall m \in \mathcal{M}, \hspace{1mm}\forall n \in \mathcal{N},
\end{align}
where $t_{k,m,n}$ is the $((m-1)N+n)$-th element of $\mathbf{t}_k$. Then, we can rewrite \eqref{t_1} as follows 
\begin{align} \label{t_2}
r_{k,m,n} t_{k,m,n} = \beta e^{j \theta_{k,m,n}}, \hspace{1mm} \forall k \in \mathcal{K}, \hspace{1mm}\forall m \in \mathcal{M}, \hspace{1mm}\forall n \in \mathcal{N},
\end{align}
A compact version of \eqref{t_2} is
\begin{align}
\mbox{C7:} \hspace{1mm} \mathbf{R}_k \mathbf{t}_k = \mathbf{u}_k, \hspace{1mm}\forall k \in \mathcal{K},
\end{align}
where $\mathbf{R}_k \in \mathbb{C}^{MN \times MN}$ is a diagonal matrix, whose $((m-1)N+n)$-th main diagonal element is $r_{k,m,n}$. $\mathbf{u}_k \in \mathbb{C}^{MN \times 1}$ is a vector whose $((m-1)N+n)$-th element is $\beta e^{j \theta_{k,m,n}}$.
\par
Based on \eqref{vectort_k}, constraint C1 can be reformulated equivalently as follows
\begin{align}
\overline{\mbox{C1}} \mbox{:} \hspace{1mm} \frac{\left| \mathbf{t}_k^H \mathbf{A} \mathbf{w}_k \right|^2}{ \underset{k' \in \mathcal{K}\setminus \{k\}}{\sum}\left| \mathbf{t}_k^H \mathbf{A} \mathbf{w}_{k'} \right|^2 + \sigma_k^2} \geq \Gamma_k, \hspace{1mm} \forall k \in \mathcal{K}.
\end{align}
Then, problem \eqref{continuous} is recast as 
\begin{align} \label{prob_admm}
\underset{\substack{ \mathbf{w}_k, \bm{\alpha}_m, \mathbf{X}, \\ \bm{\theta}, \mathbf{t}_k }}{\mino} \hspace*{3mm} & f_1 + \mathbb{I}_{\mathcal{F}}(\mathbf{w}_k, \bm{\alpha}_m, \mathbf{X},  \mathbf{t}_k) \notag\\
\subto\hspace*{3mm} & \mbox{C6:}\hspace*{1mm} \bm{\theta} = \bm{\phi}, \notag \\
& \mbox{C7:}\hspace*{1mm} \mathbf{R}_k \mathbf{t}_k = \mathbf{u}_k, \forall k \in \mathcal{K},
\end{align}
where $\mathbb{I}_{\mathcal{F}}(\mathbf{w}_k, \bm{\alpha}_m, \mathbf{X},  \mathbf{t}_k)$ is an indicator function defined as 
\begin{align}
\mathbb{I}_{\mathcal{F}}(\mathbf{w}_k, \bm{\alpha}_m, \mathbf{X},  \mathbf{t}_k) \overset{\triangle}{=}
\begin{cases}
0 & \hspace{-3mm}: (\mathbf{w}_k, \bm{\alpha}_m, \mathbf{X},  \mathbf{t}_k) \in \mathcal{F}, \\
\infty & \hspace{-3mm}: \text{otherwise},
\end{cases}
\end{align}
where $\mathcal{F}$ is the feasible set defined by constraints $\overline{\mbox{C1}}$, C2, C3, C4, and C5 and is given by
\begin{align}
\mathcal{F} \overset{\triangle}{=} \left\{  (\mathbf{w}_k, \bm{\alpha}_m, \mathbf{X},  \mathbf{t}_k) \ | \ \overline{\mbox{C1}}, \mbox{C2-C5} \right\}.
\end{align}
\par
Then, to handle equality constraints C6 and C7 and variable coupling, we apply the principle of ADMM to optimization problem \eqref{prob_admm}.
In particular, a canonical form of the augmented Lagrangian function of problem \eqref{prob_admm} is given by
\begin{align} \label{lagrange_1}
& \mathcal{L}(\mathbf{w}_k,\bm{\alpha}_m,\mathbf{X}, \bm{\theta}, \mathbf{t}_k, \bm{\mu}, \bm{\lambda}_k) \notag \\ 
= & f_1 + \mathbb{I}_{\mathcal{F}}(\mathbf{w}_k, \bm{\alpha}_m, \mathbf{X},  \mathbf{t}_k) + \frac{\rho_1}{2} \left\| \bm{\theta} - \bm{\phi} + \frac{\bm{\mu}}{\rho_1} \right\|^2_2 \notag \\
& \hspace{-3mm} + \frac{\rho_2}{2} \underset{k \in \mathcal{K}}{\sum} \left\| \mathbf{R}_k \mathbf{t}_k \hspace{-1mm} - \hspace{-1mm} \mathbf{u}_k \hspace{-1mm} + \hspace{-1mm} \frac{\bm{\lambda}_k}{\rho_2} \right\|^2_2 \hspace{-1mm} - \hspace{-1mm} \frac{\rho_1}{2} \left\| \frac{\bm{\mu}}{\rho_1} \right\|^2_2 \hspace{-1mm} - \hspace{-1mm} \frac{\rho_2}{2} \underset{k \in \mathcal{K}}{\sum} \left\| \frac{\bm{\lambda}_k}{\rho_2} \right\|^2_2,
\end{align}
where $\bm{\mu} \in \mathbb{R}^{KMN \times 1}$ and $\bm{\lambda}_k \in \mathbb{C}^{MN \times 1}$ are the dual variables with respect to equality constraints C6 and C7, respectively. Constants $\rho_1$, $\rho_2 > 0$ are penalty parameters.
The primal variables can be obtained by alternatingly minimizing the augmented Lagrangian function $\mathcal{L}(\mathbf{w}_k,\bm{\alpha}_m,\mathbf{X}, \bm{\theta}, \mathbf{t}_k, \bm{\mu}, \bm{\lambda}_k)$ with respect to $\mathbf{w}_k$, $\bm{\alpha}_m$, $\mathbf{X}$, $\bm{\theta}$, and $\mathbf{t}_k$. Given the updated primal variables in the $i$-th iteration of ADMM, the dual variables are updated as 
\begin{align}
& \bm{\mu}^{(i)} = \bm{\mu}^{(i-1)} + \rho_1 \left( \bm{\theta}^{(i)} - \bm{\phi}^{(i)} \right), \label{update_mu} \\
& \bm{\lambda}_k^{(i)} = \bm{\lambda}_k^{(i-1)} + \rho_2 \left(\mathbf{R}_k^{(i)} \mathbf{t}_k^{(i)} - \mathbf{u}_k^{(i)} \right). \label{update_lambda}
\end{align}
\par
So far, we have established the basic framework of ADMM. Next, we minimize $\mathcal{L}(\mathbf{w}_k,\bm{\alpha}_m,\mathbf{X}, \bm{\theta}, \mathbf{t}_k, \bm{\mu}, \bm{\lambda}_k)$ by alternatingly updating the primal variables $\mathbf{w}_k$, $\bm{\alpha}_m$, $\mathbf{X}$, $\bm{\theta}$, and $\mathbf{t}_k$.
\subsection{Sub-problem for Updating $\mathbf{w}_k$}
By fixing the other optimization variables, the transmit beamforming vector $\mathbf{w}_k$ is updated by solving
\begin{align}
\underset{\substack{ \mathbf{w}_k }}{\mino} \hspace*{3mm} \frac{T_2}{T_1 + T_2} \underset{k \in \mathcal{K}}{\sum} \| \mathbf{w}_k\|^2_2 + \mathbb{I}_{\mathcal{F}}(\mathbf{w}_k, \bm{\alpha}_m, \mathbf{X},  \mathbf{t}_k),
\end{align}
which is equivalent to
\begin{align} \label{block_W}
\underset{\substack{ \mathbf{w}_k }}{\mino} \hspace*{3mm} & \frac{T_2}{T_1 + T_2} \underset{k \in \mathcal{K}}{\sum} \| \mathbf{w}_k\|^2_2 \notag\\
\subto\hspace*{3mm} & \overline{\mbox{C1}} \mbox{:} \hspace{1mm} \gamma_k \geq \Gamma_k, \hspace{1mm} \forall k \in \mathcal{K}.
\end{align}
Note that any phase rotated version of $\mathbf{w}_k$ satisfies constraint $\overline{\mbox{C1}}$ while the value of the objective function remains unchanged. Hence, the term $\mathbf{t}_k^H \mathbf{A} \mathbf{w}_k$ can be chosen to be real-valued without loss of generality. As a result, the above optimization problem can be equivalently transformed as follows
\begin{align} \label{socp}
\underset{\substack{ \mathbf{w}_k, a }}{\mino} \hspace*{2mm} & a \notag\\
\subto\hspace*{2mm} & \overline{\mbox{C1a}}\mbox{:} \left\lVert 
\begin{array}{c}
\hspace{-1mm} \mathbf{W}^H \mathbf{A}^H \mathbf{t}_k \\ \sigma_k 
\end{array}
\right\rVert \hspace{-1mm} \leq \hspace{-1mm} \sqrt{1\hspace{-1mm} + \hspace{-1mm} \frac{1}{\Gamma_k}} \Re\left\{\mathbf{t}_k^H \mathbf{A} \mathbf{w}_k\right\}, 
\notag \\ & \hspace{8mm} \forall k \in \mathcal{K}, \notag \\
& \overline{\mbox{C1b}} \mbox{:} \hspace{1mm} \Im\left\{ \mathbf{t}_k^H \mathbf{A} \mathbf{w}_k \right\} = 0, \hspace{1mm} \forall k \in \mathcal{K}, \notag \\
& \mbox{C8:} \hspace{1mm} \| \mathbf{W} \|^2_2 \leq \frac{(T_1+T_2)a}{T_2},
\end{align}
where $\mathbf{W} \overset{\triangle}{=} [\mathbf{w}_1, \dots, \mathbf{w}_K]$. It can be verified that problem \eqref{socp} is a standard second-order cone programming (SOCP) problem, which can be efficiently solved by standard convex optimization solvers such as CVX \cite{boyd2004convex, grant2014cvx}.
\par

\subsection{Sub-problem for Updating $\mathbf{t}_k$}
By fixing the other optimization variables, the auxiliary variable $\mathbf{t}_k$ is updated by solving
\begin{align} \label{t_lag}
\underset{\substack{ \mathbf{t}_k }}{\mino} \frac{\rho_2}{2} \hspace{-1mm} \underset{k \in \mathcal{K}}{\sum} \left\| \mathbf{R}_k \mathbf{t}_k \hspace{-1mm} - \hspace{-1mm} \mathbf{u}_k \hspace{-1mm} + \hspace{-1mm} \frac{\bm{\lambda}_k}{\rho_2} \right\|^2_2 \hspace{-1mm} + \hspace{-1mm} \mathbb{I}_{\mathcal{F}}(\mathbf{w}_k, \bm{\alpha}_m, \mathbf{X},  \mathbf{t}_k),
\end{align}
which can be equivalently rewritten as follows
\begin{align} \label{t_k}
\underset{\substack{ \mathbf{t}_k }}{\mino} \hspace*{3mm} & \frac{\rho_2}{2} \underset{k \in \mathcal{K}}{\sum} \left\| \mathbf{R}_k \mathbf{t}_k - \mathbf{u}_k + \frac{\bm{\lambda}_k}{\rho_2} \right\|^2_2 \notag\\
\subto\hspace*{2mm} & \overline{\mbox{C1}} \mbox{:} \hspace{1mm} \gamma_k \geq \Gamma_k, \hspace{1mm} \forall k \in \mathcal{K}.
\end{align}
Different from the problem with respect to $\mathbf{w}_k$ in \eqref{block_W}, a phase rotation of $\mathbf{t}_k$ has an impact on the objective function value. As a result, the transformation used in \eqref{socp} cannot be applied to problem \eqref{t_k}. Instead, we first rewrite $\overline{\mbox{C1}}$ equivalently as
\begin{align} \label{c1_}
\hspace{-2mm}\overline{\mbox{C1}} \mbox{:} \hspace{1mm} \Gamma_k  \hspace{-1mm} \underset{k' \in \mathcal{K}\setminus \{k\}}{\sum} \hspace{-1mm} \left| \mathbf{t}_k^H \mathbf{A} \mathbf{w}_{k'} \right|^2 - \left| \mathbf{t}_k^H \mathbf{A} \mathbf{w}_k \right|^2  + \sigma_k^2 \Gamma_k \leq 0, \hspace{1mm} \forall k,
\end{align}
where the left-hand side of the inequality is a difference of convex (d.c.) functions with respect to $\mathbf{t}_k$ which can be effectively tackled by SCA. Specifically, we construct an upper-bound on the non-convex term $-\left| \mathbf{t}_k^H \mathbf{A} \mathbf{w}_k \right|^2$ by employing its first-order Taylor expansion. 
Constraint $\overline{\mbox{C1}}$ is transformed into
\begin{align}
\myoverline{\mbox{C1}} \mbox{:} \hspace{1mm} & \Gamma_k \hspace{-1mm} \underset{k' \in \mathcal{K}\setminus \{k\}}{\sum} \hspace{-1mm} \left| \mathbf{t}_k^H \mathbf{A} \mathbf{w}_{k'} \right|^2 -2 \Re\left\{ \left(\mathbf{t}^{(i-1)}_k\right)^H \mathbf{A} \mathbf{w}_k \mathbf{w}_k^H \mathbf{A}^H \mathbf{t}_k \right\} \notag \\
& + \left| \left(\mathbf{t}^{(i-1)}_k\right)^H \mathbf{A} \mathbf{w}_k \right|^2 + \sigma_k^2 \Gamma_k \leq 0, \hspace{1mm} \forall k \in \mathcal{K},
\end{align}
where $\mathbf{t}^{(i-1)}_k$ denotes the solution of $\mathbf{t}_k$ obtained in the previous iteration.
As a result, problem \eqref{t_k} is replaced by the following approximated optimization problem 
\begin{align} \label{sca}
\underset{\substack{ \mathbf{t}_k }}{\mino} \hspace*{3mm} & \frac{\rho_2}{2} \underset{k \in \mathcal{K}}{\sum} \left\| \mathbf{R}_k \mathbf{t}_k - \mathbf{u}_k + \frac{\bm{\lambda}_k}{\rho_2} \right\|^2_2 \notag\\
\subto\hspace*{2mm} & \myoverline{\mbox{C1}},
\end{align}
which is convex and can be solved efficiently.
\subsection{Sub-problem for Updating $\bm{\alpha}_m$}
By fixing the other optimization variables, the antenna radiation power ratios $\bm{\alpha}_m$ are updated by solving
\begin{align}
\underset{\substack{ \bm{\alpha}_m }}{\mino} \hspace*{3mm} \mathbb{I}_{\mathcal{F}}(\mathbf{w}_k, \bm{\alpha}_m, \mathbf{X},  \mathbf{t}_k),
\end{align}
which is equivalent to the following feasibility check problem
\begin{align} \label{alpha_feasibility_continuous}
\underset{\substack{ }}{\mathrm{Find}} \hspace*{3mm} & \bm{\alpha}_m   \notag\\
\subto\hspace*{3mm} & \overline{\mbox{C1}}, \mbox{C2}.
\end{align}
To accelerate the convergence and to provide more DoFs for energy minimization in other sub-problems, we update $\bm{\alpha}_m$ by maximizing the SINR among users \cite{10364735}, which leads to the following problem
\begin{align} \label{alpha_1}
\underset{\substack{ \bm{\alpha}_m }}{\maxo} \hspace*{3mm} & \underset{k \in \mathcal{K}}{\sum} \frac{\left| \mathbf{t}_k^H \mathbf{A} \mathbf{w}_k \right|^2}{ \underset{k' \in \mathcal{K}\setminus \{k\}}{\sum}\left| \mathbf{t}_k^H \mathbf{A} \mathbf{w}_{k'} \right|^2 + \sigma_k^2} \notag\\
\subto\hspace*{2mm} & \mbox{C2:} \hspace{1mm}  \left\| \bm{\alpha}_m \right\|^2_2 \leq 1, \hspace{1mm} \forall m \in \mathcal{M},
\end{align}
where the non-convexity is caused by the fractional expression. To deal with this issue, fractional programming is exploited. By introducing auxiliary variable $\mathbf{q} = [q_1, \dots, q_K]^T \in \mathbb{R}^{K \times 1}$, problem \eqref{alpha_1} is then equivalently transformed into \cite{8314727}
\begin{align} \label{fp}
\underset{\substack{ \bm{\alpha}_m, \mathbf{q} }}{\maxo} \hspace*{3mm} & f_2 \overset{\triangle}{=} \underset{k \in \mathcal{K}}{\sum} \Bigg( 2 q_k \left| \mathbf{t}_k^H \mathbf{A} \mathbf{w}_k \right| \notag \\
& - q_k^2 \bigg( \underset{k' \in \mathcal{K}\setminus \{k\}}{\sum}\left| \mathbf{t}_k^H \mathbf{A} \mathbf{w}_{k'} \right|^2 + \sigma_k^2 \bigg) \Bigg) \notag\\
\subto\hspace*{3mm} & \mbox{C2:} \hspace{1mm}  \left\| \bm{\alpha}_m \right\|^2_2 \leq 1, \hspace{1mm} \forall m \in \mathcal{M}.
\end{align}
For given $\bm{\alpha}_m$, the optimal $\mathbf{q}$ is obtained from $\frac{\partial f_2}{\partial q_k} = 0, \forall k \in \mathcal{K}$, given by
\begin{align} \label{cal_q_1}
q_k = \frac{\left| \mathbf{t}_k^H \mathbf{A} \mathbf{w}_k \right| }{\underset{k' \in \mathcal{K}\setminus \{k\}}{\sum}\left| \mathbf{t}_k^H \mathbf{A} \mathbf{w}_{k'} \right|^2 + \sigma_k^2}, \hspace{1mm} \forall k \in \mathcal{K}.
\end{align}
Next, we solve problem \eqref{fp} for given $\mathbf{q}$. To begin with, we introduce auxiliary variable $\bm{\psi} \in \mathbb{R}^{K \times 1}$ whose the $k$-th element is $\psi_k$ and the following constraint
\begin{align}
\mbox{C9:} \hspace{1mm} \left| \mathbf{t}_k^H \mathbf{A} \mathbf{w}_k \right| \geq \psi_k, \hspace{1mm} \forall k \in \mathcal{K}.
\end{align}
The resulting objective function is then expressed as
\begin{align}
\overline{f}_2 \overset{\triangle}{=} \underset{k \in \mathcal{K}}{\sum} \left( 2 q_k \psi_k - q_k^2 \left( \underset{k' \in \mathcal{K}\setminus \{k\}}{\sum}\left| \mathbf{t}_k^H \mathbf{A} \mathbf{w}_{k'} \right|^2 + \sigma_k^2 \right) \right).
\end{align}
Note that constraint C9 is non-convex with respect to $\bm{\alpha}_m$. To handle this,
we first equivalently transform C9 into 
\begin{align}
\mbox{C9:} \hspace{1mm} \mathbf{t}_k^H \mathbf{A} \mathbf{w}_k \mathbf{w}_k^H \mathbf{A}^T \mathbf{t}_k \geq \psi_k^2,  \hspace{1mm} \forall k \in \mathcal{K}.
\end{align}
By constructing a lower bound on the term $\mathbf{t}_k^H \mathbf{A} \mathbf{w}_k \mathbf{w}_k^H \mathbf{A}^T \mathbf{t}_k$, 
constraint C9 can be rewritten as
\begin{align}
& \overline{\mbox{C9}} \mbox{:} \hspace{1mm} \mathrm{Tr}\left( \left( \mathbf{T}_k \mathbf{A}^{(i-1)} \mathbf{W}_k + \mathbf{T}_k^T \mathbf{A}^{(i-1)} \mathbf{W}_k^T \right)^T \mathbf{A} \right) \notag \\
& - \mathrm{Tr} \left( \mathbf{T}_k \mathbf{A}^{(i-1)} \mathbf{W}_k \left(\mathbf{A}^{(i-1)}\right)^T \right) - \psi_k^2 \geq 0, \hspace{1mm} \forall k \in \mathcal{K},
\end{align}
where $\mathbf{T}_k \overset{\triangle}{=} \mathbf{t}_k \mathbf{t}_k^H$, $\mathbf{W}_k \overset{\triangle}{=} \mathbf{w}_k \mathbf{w}_k^H$. $\mathbf{A}^{(i-1)}$ is the solution obtained in the previous iteration. 
With $\mathbf{q}$ obtained from \eqref{cal_q_1}, optimization problem \eqref{fp} for solving $\bm{\alpha}_m$ is then transformed into
\begin{align} \label{prob_alpha_2}
\underset{\substack{ \bm{\alpha}_m, \bm{\psi} }}{\maxo} \hspace*{3mm} & \overline{f}_2 \notag\\
\subto\hspace*{3mm} & \mbox{C2}, \overline{\mbox{C9}},
\end{align}
which is convex and can be solved efficiently.

\begin{figure*}[b]
\hrule
\setcounter{equation}{59}
\begin{align} \label{opt_theta}
\theta_{k,m,n} = \frac{2\rho_1(\phi_{k,m,n}-\frac{\mu_{k,m,n}}{\rho_1}) + \rho_2 L_{k,m,n} \theta_{k,m,n}^{(i-1)} - \rho_2 \nabla_{\theta} v_{k,m,n}\left(\theta_{k,m,n}^{(i-1)}\right)}{2\rho_1 + \rho_2 L_{k,m,n}}.
\end{align}
\setcounter{equation}{51}
\end{figure*}

\par
\subsection{Sub-problem for Updating $\bm{\theta}$}
The auxiliary variable $\bm{\theta}$ is updated by solving
\begin{align} \label{prob_theta}
\underset{\substack{ \bm{\theta} }}{\mino} \hspace*{1mm} & f_3 \overset{\triangle}{=} \frac{\rho_1}{2} \left\| \bm{\theta} - \bm{\phi} + \frac{\bm{\mu}}{\rho_1} \right\|^2_2\hspace*{-2mm}+ \frac{\rho_2}{2} \hspace*{-1mm} \underset{k \in \mathcal{K}}{\sum} \left\| \mathbf{R}_k \mathbf{t}_k \hspace*{-1mm} - \hspace*{-1mm} \mathbf{u}_k  \hspace*{-1mm} + \hspace*{-1mm} \frac{\bm{\lambda}_k}{\rho_2} \right\|^2_2.
\end{align}
Here, the non-convexity comes from the second term, which can be explicitly expressed as $\frac{\rho_2}{2} \underset{\substack{k, m, n}}{\sum} v_{k,m,n}(\theta_{k,m,n})$, where $v_{k,m,n}(\theta_{k,m,n})$ is defined as
\begin{equation}
v_{k,m,n}(\theta_{k,m,n})\overset{\triangle}{=} \left| r_{k,m,n} t_{k,m,n} \hspace{-0.5mm} - \hspace{-0.5mm} \beta e^{j \theta_{k,m,n}} \hspace{-0.5mm} + \hspace{-0.5mm} \frac{\lambda_{k,m,n}}{\rho_2} \right|^2.
\end{equation}
The variable $\lambda_{k,m,n}$ is the $((m-1)N + n)$-th element of $\bm{\lambda}_k$. We note that the variable $\theta_{k,m,n}$ is involved in the exponential term, which is neither convex nor concave. To handle this challenge, we construct an upper-bound on $v_{k,m,n}(\theta_{k,m,n})$ based on the following lemma \cite{7547360, xu2025joint}.
\begin{lemma}
\textit{(Descent Lemma \cite{bertsekas1997nonlinear})}: Let $f: \mathbb{R}^n \rightarrow \mathbb{R}$ be a continuously differentiable function with a Lipschitz continuous gradient on $\mathcal{X} \in \mathbb{R}^n$ with Lipschitz constant $L$. Then, for all $\mathbf{x}, \mathbf{y} \in \mathcal{X}$, we have
\begin{align}
f(\mathbf{x}) \leq f(\mathbf{y}) + \nabla f(\mathbf{y})^T(\mathbf{x} - \mathbf{y}) + \frac{L}{2} \| \mathbf{x} - \mathbf{y} \|^2_2.
\end{align}
\end{lemma}

\begin{Prop}
$v_{k,m,n}(\theta_{k,m,n})$ has a Lipschitz continuous gradient on $\mathbb{R}$ with the Lipschitz constant $L_{k,m,n}$ given by
\begin{align}
L_{k,m,n} = 2 \beta \left|r_{k,m,n} t_{k,m,n} + \frac{\lambda_{k,m,n}}{\rho_2} \right|.
\end{align}
\end{Prop}
\textit{Proof:} Please refer to the Appendix.
\QED
\par
Based on Proposition 1, an upper-bound on $v_{k,m,n}(\theta_{k,m,n})$ is constructed as follows
\begin{align}
& v_{k,m,n}(\theta_{k,m,n}) \notag \\
& \leq v_{k,m,n}\left(\theta_{k,m,n}^{(i-1)}\right) + \nabla_{\theta} v_{k,m,n}\left(\theta_{k,m,n}^{(i-1)}\right) \left(\theta_{k,m,n}- \theta_{k,m,n}^{(i-1)}\right) \notag \\
& \hspace{4mm} + \frac{L_{k,m,n}}{2} \left( \theta_{k,m,n}- \theta_{k,m,n}^{(i-1)} \right)^2 \notag \\
& \overset{\triangle}{=} \widetilde{v}_{k,m,n}(\theta_{k,m,n}),
\end{align}
where 
\begin{align}
&\nabla_{\theta} v_{k,m,n}\left(\theta_{k,m,n}^{(i-1)}\right) = 2 \beta \left|r_{k,m,n} t_{k,m,n} + \frac{\lambda_{k,m,n}}{\rho_2} \right| \notag \\ \times 
& \sin \left( \theta_{k,m,n}^{(i-1)} - \angle \left( r_{k,m,n}t_{k,m,n}+ \frac{\lambda_{k,m,n}}{\rho_2} \right) \right).
\end{align}
Here, $\theta_{k,m,n}^{(i-1)}$ denotes the solution of $\theta_{k,m,n}$ obtained in the previous iteration.
As a result, the second-order surrogate function of objective function $f_3$ in \eqref{prob_theta} is given by
\begin{align}
f_3 & \leq \hspace*{-1.5mm}\underset{\substack{k, m, n}}{\sum}\hspace*{-2mm}\left( \hspace*{-1mm} \frac{\rho_1}{2} \hspace*{-1mm} \left( \theta_{k,m,n}\hspace*{-0.5mm}-\hspace*{-0.5mm} \phi_{k,m,n}\hspace*{-0.5mm}+\hspace*{-0.5mm}\frac{\mu_{k,m,n}}{\rho_1} \right)^2 \hspace*{-2mm}+ \hspace*{-1mm}\frac{\rho_2}{2} \widetilde{v}_{k,m,n}(\theta_{k,m,n}) \hspace*{-1mm} \right) \notag \\
& \overset{\triangle}{=} \widetilde{f}_3,
\end{align}
where $\mu_{k,m,n}$ denotes the $((k-1)MN + (m-1)N + n)$-th element of $\bm{\mu}$.
Problem \eqref{prob_theta} is transformed into
\begin{align} \label{theta_2}
\underset{\substack{ \bm{\theta} }}{\mino} \hspace*{3mm} & \widetilde{f}_3,
\end{align}
which has the closed-form solution given by \eqref{opt_theta}, shown at the bottom of the page.
\setcounter{equation}{60}

\par
\subsection{Sub-problem for Updating $\mathbf{X}$}
The antenna positions $\mathbf{X}$ are updated by solving
\begin{align} \label{x_prob}
\underset{\substack{ \mathbf{X} }}{\mino} \hspace*{3mm} & \frac{1}{T_1 + T_2} \underset{\substack{m, n}}{\sum} E_{m,n} + \mathbb{I}_{\mathcal{F}}(\mathbf{w}_k, \bm{\alpha}_m, \mathbf{X},  \mathbf{t}_k) \notag \\ & \hspace{-15mm} + \frac{\rho_1}{2} \left\| \bm{\theta} - \bm{\phi} + \frac{\bm{\mu}}{\rho_1} \right\|^2_2  + \frac{\rho_2}{2} \underset{k \in \mathcal{K}}{\sum} \left\| \mathbf{R}_k \mathbf{t}_k - \mathbf{u}_k + \frac{\bm{\lambda}_k}{\rho_2} \right\|^2_2,
\end{align}
which is equivalent to
\begin{align} \label{optimize_X}
\underset{\substack{ \mathbf{X} }}{\mino} \hspace*{3mm} & f_4 \overset{\triangle}{=} \frac{1}{T_1 + T_2} \underset{\substack{m, n}}{\sum} E_{m,n} + \frac{\rho_1}{2} \left\| \bm{\theta} - \bm{\phi} + \frac{\bm{\mu}}{\rho_1} \right\|^2_2 \notag \\
&  \hspace{8mm}  + \frac{\rho_2}{2} \underset{k \in \mathcal{K}}{\sum} \left\| \mathbf{R}_k \mathbf{t}_k - \mathbf{u}_k + \frac{\bm{\lambda}_k}{\rho_2} \right\|^2_2\notag\\
\subto\hspace*{2mm} & \mbox{C3-C5}.
\end{align}
By substituting the locations of the antennas and users, the objective function is formulated as
\begin{align} \label{f_4}
& f_4 = \frac{P}{(T_1 + T_2)v} \underset{\substack{m, n}}{\sum} |x_{m,n} - x'_{m,n}| \notag \\
& + \frac{\rho_1}{2} \underset{\substack{k, m, n}}{\sum} \left( \theta_{k,m,n} - \frac{2 \pi}{\lambda_c} ( r_{k,m,n}+n_{\text{eff}}x_{m,n}) + \frac{\mu_{k,m,n}}{\rho_1} \right)^2 \notag \\
& + \frac{\rho_2}{2} \underset{\substack{k, m, n}}{\sum} \left| r_{k,m,n} t_{k,m,n} - \beta e^{j\theta_{k,m,n}} + \frac{\lambda_{k,m,n}}{\rho_2} \right|^2.
\end{align}
\par
\par
Here, the non-convexity of the objective function comes from the last two terms. To deal with the second term, we introduce auxiliary variables $\xi_{k,m,n} \geq 0, \hspace{1mm} \forall k, \forall m, \forall n$, and the following constraint
\begin{align}
\mbox{C10:} \hspace{1mm} & \left| \theta_{k,m,n} - \frac{2 \pi}{\lambda_c} (r_{k,m,n}+n_{\text{eff}}x_{m,n}) + \frac{\mu_{k,m,n}}{\rho_1} \right| \leq \xi_{k,m,n}, \notag \\
 & \forall k, \forall m, \forall n,
\end{align}
which is equivalent to
\begin{align}
& \mbox{C10a:} \hspace{1mm} \theta_{k,m,n} - \frac{2 \pi}{\lambda_c} (r_{k,m,n}+n_{\text{eff}}x_{m,n}) + \frac{\mu_{k,m,n}}{\rho_1} \geq -\xi_{k,m,n}, \notag \\ 
& \hspace{10mm} \forall k \in \mathcal{K}, \forall m \in \mathcal{M}, \forall n \in \mathcal{N}, \\
& \mbox{C10b:} \hspace{1mm} \theta_{k,m,n} - \frac{2 \pi}{\lambda_c} (r_{k,m,n}+n_{\text{eff}}x_{m,n}) + \frac{\mu_{k,m,n}}{\rho_1} \leq \xi_{k,m,n}, \notag \\
& \hspace{10mm} \forall k \in \mathcal{K}, \forall m \in \mathcal{M}, \forall n \in \mathcal{N},
\end{align}
where constraint C10a is convex but C10b is non-convex due to the term $r_{k,m,n}$. To tackle this issue, we construct a lower bound with respect to $r_{k,m,n}$, which is given by
\begin{align} \label{linear_r}
r_{k,m,n} & \geq r_{k,m,n}^{(i-1)} + \frac{x_{m,n}^{(i-1)} - \overline{x}_k}{r_{k,m,n}^{(i-1)}} \left( x_{m,n} - x_{m,n}^{(i-1)} \right) \notag \\
& \overset{\triangle}{=} \widetilde{r}_{k,m,n},
\end{align}
where $r_{k,m,n}^{(i-1)}$ denotes the value of $r_{k,m,n}$ evaluated using $x_{m,n}^{(i-1)}$ obtained in the previous iteration. Constraint C10b is then transformed into
\begin{align}
\overline{\mbox{C10b}} \mbox{:} \hspace{1mm} & \frac{2 \pi}{\lambda_c} (\widetilde{r}_{k,m,n}+n_{\text{eff}}x_{m,n}) +\xi_{k,m,n}\geq \theta_{k,m,n} + \frac{\mu_{k,m,n}}{\rho_1}, \notag \\
& \forall k, \forall m, \forall n.
\end{align}
\par
Next, we handle the term $\frac{\rho_2}{2} \underset{k \in \mathcal{K}}{\sum} \left\| \mathbf{R}_k \mathbf{t}_k - \mathbf{u}_k + \frac{\bm{\lambda}_k}{\rho_2} \right\|^2_2$ in objective function $f_4$ in \eqref{optimize_X}. To begin with, we define $\eta_{k,m,n}(x_{m,n}) \overset{\triangle}{=} \left| r_{k,m,n} t_{k,m,n} - \beta e^{j\theta_{k,m,n}} + \frac{\lambda_{k,m,n}}{\rho_2} \right|^2 $, which can be expanded as
\begin{align}
\eta_{k,m,n} = & |t_{k,m,n}|^2 r_{k,m,n}^2 + \left| \frac{\lambda_{k,m,n}}{\rho_2} - \beta e^{j \theta_{k,m,n}} \right|^2 \notag \\
& \hspace{-3mm} + 2 \left| \frac{\lambda_{k,m,n}}{\rho_2} - \beta e^{j \theta_{k,m,n}} \right| |t_{k,m,n}| \cos\left( \vartheta_{k,m,n} \right) r_{k,m,n},
\end{align}
where $\vartheta_{k,m,n} \overset{\triangle}{=} \angle t_{k,m,n} - \angle\left( \frac{\lambda_{k,m,n}}{\rho_2} - \beta e^{j\theta_{k,m,n}} \right)$. The convexity of $\eta_{k,m,n}$ depends on the value of $\cos\left( \vartheta_{k,m,n} \right)$. Specifically, $\eta_{k,m,n}$ is convex if $\cos\left( \vartheta_{k,m,n} \right) \geq 0$. Otherwise, $\eta_{k,m,n}$ is non-convex and is in the form of a d.c. function. For handling the non-convex case of $\vartheta_{k,m,n}$, we linearize $r_{k,m,n}$ by $\widetilde{r}_{k,m,n}$ as in \eqref{linear_r}, leading to a convex upper bound on $\eta_{k,m,n}$, denoted by $\widetilde{\eta}_{k,m,n}$. As a result, the surrogate function of the term $\left| r_{k,m,n} t_{k,m,n} - \beta e^{j\theta_{k,m,n}} + \frac{\lambda_{k,m,n}}{\rho_2} \right|^2$ is obtained as
\begin{align}
\overline{\eta}_{k,m,n} = 
\begin{cases}
    \eta_{k,m,n}, \hspace{2mm} \cos\left( \vartheta_{k,m,n} \right) \geq 0, \\
    \widetilde{\eta}_{k,m,n}, \hspace{2mm} \text{otherwise}.
\end{cases}
\end{align}
We note that the term $\vartheta_{k,m,n}$ is a constant when optimizing $\mathbf{X}$, hence the expression for $\overline{\eta}_{k,m,n}$ is known before solving problem \eqref{optimize_X}.
The objective function is now formulated as follows
\begin{align}
\overline{f}_4 = & \frac{P}{(T_1 + T_2)v} \hspace*{-1mm} \underset{\substack{m, n}}{\sum} |x_{m,n} \hspace*{-1mm} - \hspace*{-1mm} x'_{m,n}| \hspace*{-1mm} + \hspace*{-3mm} \underset{\substack{k, m, n}}{\sum} \hspace*{-2mm} \left( \frac{\rho_1}{2} \xi_{k,m,n}^2 \hspace*{-1mm} + \hspace*{-1mm} \frac{\rho_2}{2} \overline{\eta}_{k,m,n} \right).
\end{align}
Thus, optimization problem \eqref{optimize_X} is transformed into
\begin{align} \label{prob_x_2}
\underset{\substack{ \mathbf{X}, \bm{\xi} }}{\mino} \hspace*{4mm} & \overline{f}_4 \notag\\
\subto\hspace*{2mm} & \mbox{C3-C5}, \mbox{C10a}, \overline{\mbox{C10b}},
\end{align}
which is a convex optimization problem and can be solved efficiently. 
\par
The ADMM-based algorithm for solving problem \eqref{continuous} for continuous antenna movement is summarized in $\textbf{Algorithm 1}$, where we set the scaling factor $\epsilon > 1$ to gradually enlarge the penalty parameters to speed up the elimination of constraint violations. and force the equality constraints to be satisfied. $f_1^{(i)}$ denotes the value of $f_1$ obtained in the $i$-th iteration. The convergence and computational complexity of Algorithm 1 are discussed in the following.
\par
The objective function values of problems \eqref{sca}, \eqref{theta_2}, \eqref{prob_x_2} serve as upper bounds on the objective function values of problems \eqref{t_lag}, \eqref{prob_theta}, \eqref{x_prob}, respectively. For any fixed penalty parameters, these upper bounds are monotonically tightened by alternatingly solving \eqref{sca}, \eqref{theta_2}, and \eqref{prob_x_2} for updating $\mathbf{t}_k$, $\bm{\theta}$, and $\mathbf{X}$, respectively. Together by alternatingly solving \eqref{socp} and \eqref{prob_alpha_2} for updating $\mathbf{w}_k$ and $\bm{\alpha}_m$, the augmented Lagrangian function $\mathcal{L}(\mathbf{w}_k,\bm{\alpha}_m,\mathbf{X}, \bm{\theta}, \mathbf{t}_k, \bm{\mu}, \bm{\lambda}_k)$ is non-increasing and the corresponding solution is guaranteed to converge to a stationary point of problem \eqref{continuous} \cite{boyd2011distributed}.
%
%
%
The computational complexity of Algorithm 1 is dominated by updating variables $\mathbf{t}_k$, $\bm{\alpha}_m$, and $\mathbf{X}$. Updating $\mathbf{t}_k$ involves solving a quadratic constrained quadratic programming
(QCQP) problem \eqref{sca} with complexity given by $B_1 = \mathcal{O}(M^3N^3K^{3.5} + MNK^{4.5})$ \cite{ben2001lectures}. Updating $\bm{\alpha}_m$ requires solving the QCQP \eqref{prob_alpha_2} with complexity given by $B_2 = \mathcal{O}((K+M)^{0.5}M^3N^3 + ((K+M)^{0.5}MNK))$. Updating $\mathbf{X}$ requires solving SOCP problem \eqref{prob_x_2} with complexity given by $B_3 = \mathcal{O}(K^{0.5}M^{3.5}N^{3.5} + K^{1.5}M^{2.5}N^{2.5})$. As a result, the overall complexity of Algorithm 1 is given by $B_1 + B_2 + B_3$.


\begin{algorithm}[t]
\caption{ADMM-based Algorithm for Solving Problem \eqref{continuous} for Continuous Antenna Movement}
\begin{algorithmic}[1]
\small
\STATE Set iteration index $i=1$, scaling factor for the penalty parameters $\epsilon > 1$, error tolerances $\varsigma_1$ and $\varsigma_2$. Initialize $\mathbf{X}^{(1)}$ within the maximum movement distance; initialize $\bm{\theta}^{(1)}$ and $\mathbf{t}^{(1)}$ according to \eqref{C6} and \eqref{vectort_k}, respectively; initialize $\alpha_{m,n}^{(1)} = \frac{1}{\sqrt{N}}$
\REPEAT
\STATE Update $\mathbf{w}_k$ by solving problem \eqref{socp}
\STATE Update $\mathbf{t}_k$ by solving problem \eqref{sca}
\STATE Update $\bm{\alpha}_m$ by solving problem \eqref{prob_alpha_2}
\STATE Update $\bm{\theta}$ by \eqref{opt_theta}
\STATE Update $\mathbf{X}$ by solving problem \eqref{prob_x_2}
\STATE Update dual variables $\bm{\mu}$ and $\bm{\lambda}$ by \eqref{update_mu} and \eqref{update_lambda}
\STATE Update penalty parameters $\rho_1 \leftarrow \epsilon \rho_1$, $\rho_2 \leftarrow \epsilon \rho_2$
\STATE Set $i \leftarrow i + 1$
\UNTIL  $\left\| \bm{\theta} - \bm{\phi} \right\|^2_2 + \underset{k \in \mathcal{K}}{\sum} \left\| \mathbf{R}_k \mathbf{t}_k - \mathbf{u}_k \right\|^2_2 \leq \varsigma_1$ and $\frac{\left|f_1^{(i)} - f_1^{(i-1)}\right|}{f_1^{(i-1)}} \leq \varsigma_2$
\end{algorithmic}
\end{algorithm}
\section{Joint Power and Antenna Control for Discrete Antenna Movement}
In this section, we consider PASS with discrete antenna movement. Different from the case with continuous antenna movement, the possible positions of the PAs with discrete movement are discretized, leading to the joint power and antenna control problem becoming an MINLP. To handle this issue, we recast the optimization problem as an antenna position selection problem. An algorithm exploiting the BCD method and bilinear transformation is developed to tackle the formulated optimization problem.
\par
\subsection{Problem Reformulation for Discrete Antenna Movement}
\par
The discrete antenna movement problem can be seen as an antenna position selection problem. Specifically, each PA should be assigned one particular position to move toward. Assume there are $\widetilde{N}$ available discrete positions within the maximum movement distance of each PA and we define the set $\widetilde{\mathcal{N}} \overset{\triangle}{=} \left[1, \dots, \widetilde{N} \right]$ to collect the indices of the available antenna positions. The location of the $\widetilde{n}$-th available antenna position with respect to the $(m,n)$-th PA is denoted by $\widetilde{\mathbf{p}}_{m,n,\widetilde{n}} = [\widetilde{x}_{m,n,\widetilde{n}}, y_m, h]^T$, where $\widetilde{x}_{m,n,\widetilde{n}}$ denotes the corresponding $x$-axis coordinate.
To accommodate the problem in the discrete antenna movement case, we first expand the equivalent channel between the PAs and the $k$-th user $\mathbf{h}_k^H\mathbf{G} \in \mathbb{C}^{1 \times MN}$ as follows
\begin{align} \label{disc_h}
\mathbf{h}_k^H\mathbf{G} = [\overline{\mathbf{h}}_{k,1}^H, \dots, \overline{\mathbf{h}}_{k,M}^H],
\end{align}
where $\overline{\mathbf{h}}_{k,m} \in \mathbb{C}^{N \times 1}$ denotes the effective channel vector between the $m$-th waveguide and the $k$-th user. By considering the problem as a discrete antenna position selection problem, we express $\overline{\mathbf{h}}_{k,m}$ as
\begin{align} \label{h_exp}
\overline{\mathbf{h}}_{k,m}^H & = \left[\widetilde{\mathbf{h}}_{k,m,1}^H \mathbf{z}_{m,1}, \dots, \widetilde{\mathbf{h}}_{k,m,N}^H \mathbf{z}_{m,N} \right], \notag \\
& = \widetilde{\mathbf{h}}_{k,m}^H \mathbf{Z}_m,·············
\end{align}
where
$\widetilde{\mathbf{h}}_{k,m}$ and $\mathbf{Z}_m$ are defined as
\begin{align}
\widetilde{\mathbf{h}}_{k,m} & \overset{\triangle}{=} \left[ \widetilde{\mathbf{h}}_{k,m,1}^H, \dots, \widetilde{\mathbf{h}}_{k,m,N}^H \right]^H  \in \mathbb{C}^{N\widetilde{N} \times 1}, \\
\mathbf{Z}_m & \overset{\triangle}{=} \diag \left[ \mathbf{z}_{m,1}, \dots, \mathbf{z}_{m,N} \right] \in \mathbb{C}^{N\widetilde{N} \times N},
\end{align}
Here, $\widetilde{\mathbf{h}}_{k,m,n}^H \overset{\triangle}{=} [\widetilde{h}_{k,m,n,1}, \dots, \widetilde{h}_{k,m,n,\widetilde{N}}]$ denotes the channel vector from all available antenna positions of the $(m,n)$-th PA to the $k$-th user, with the elements given by
\begin{align}
\widetilde{h}_{k,m,n,\widetilde{n}} = \frac{\beta e^{-j\frac{2\pi}{\lambda_c}(\|\widetilde{\mathbf{p}}_{m,n,\widetilde{n}} - \overline{\mathbf{p}}_k \|+n_{\text{eff}}\|\widetilde{\mathbf{p}}_{m,n,\widetilde{n}} - \mathbf{p}_{m,0} \|)}}{\|\widetilde{\mathbf{p}}_{m,n,\widetilde{n}} - \overline{\mathbf{p}}_k \|}.
\end{align}
We define vector $\mathbf{z}_{m,n} \overset{\triangle}{=} [z_{m,n,1}, \dots, z_{m,n,\widetilde{N}}]^T \in \mathbb{R}^{\widetilde{N} \times 1}$ to indicate the antenna position selection for the $(m,n)$-th PA. Specifically, $z_{m,n,\widetilde{n}} = 1$ indicates that the $(m,n)$-th PA will be moved to the available antenna position $\widetilde{\mathbf{p}}_{m,n,\widetilde{n}}$, otherwise, $z_{m,n,\widetilde{n}} = 0$. One PA occupies only one position, leading to the following constraint
\begin{align}
\mbox{C11:} \hspace{1mm} \mathbf{1}_{\widetilde{N}}^T \mathbf{z}_{m,n} = 1,\forall m \in \mathcal{M}, \forall n \in \mathcal{N},
\end{align}
where $\mathbf{1}_{\widetilde{N}} \in \mathbb{R}^{\widetilde{N} \times 1}$ is a vector with all elements being equal to $1$. By substituting \eqref{h_exp} into \eqref{disc_h}, the effective channel vector $\mathbf{h}_k^H\mathbf{G}$ can be expressed as
\begin{align}
\mathbf{h}_k^H\mathbf{G} = \widetilde{\mathbf{h}}_k^H\mathbf{Z},
\end{align}
where 
\begin{align}
\widetilde{\mathbf{h}}_k & = [\widetilde{\mathbf{h}}_{k,1}^H, \dots, \widetilde{\mathbf{h}}_{k,M}^H ]^H \in \mathbb{C}^{MN\widetilde{N} \times 1}, \\
\mathbf{Z} & = \diag\left[ \mathbf{Z}_1, \dots, \mathbf{Z}_M \right]  \in \mathbb{C}^{MN\widetilde{N} \times MN}.
\end{align}
As a result, the SINR at the $k$-th user can be rewritten equivalently as
\begin{align}
\widetilde{\gamma}_k = \frac{\left| \widetilde{\mathbf{h}}_k^H\mathbf{Z} \mathbf{A} \mathbf{w}_k \right|^2}{ \underset{k' \in \mathcal{K}\setminus \{k\}}{\sum}\left| \widetilde{\mathbf{h}}_k^H\mathbf{Z} \mathbf{A} \mathbf{w}_{k'} \right|^2 + \sigma_k^2}.
\end{align}
Then, we define $\widetilde{\mathbf{x}}_{m,n} = [\widetilde{x}_{m,n,1}, \dots, \widetilde{x}_{m,n,\widetilde{N}}]^T$ to collect all available locations of the $(m,n)$-th PA along the $x$-axis, leading to $x_{m,n} = \widetilde{\mathbf{x}}_{m,n}^T \mathbf{z}_{m,n}$. Furthermore, we define the displacement vector of the $(m,n)$-th PA as 
\begin{align}
\widetilde{\mathbf{d}}_{m,n} = \left[\left| \widetilde{x}_{m,n,1} - x'_{m,n} \right|, \dots, \left| \widetilde{x}_{m,n,\widetilde{N}} - x'_{m,n} \right| \right]^T,
\end{align}
which results in $|x_{m,n} - x'_{m,n}| = \widetilde{\mathbf{d}}_{m,n}^T \mathbf{z}_{m,n}$.
\par
Based on the above transformations, the optimization problem for discrete antenna movement is given by
\begin{align} \label{discrete}
\hspace*{-3mm}\underset{\substack{ \mathbf{w}_k, \bm{\alpha}_m, \mathbf{Z} }}{\mino} \hspace*{3mm} & \hspace{-1mm} f_5 \hspace{-1mm} \overset{\triangle}{=} \hspace{-1mm} \frac{T_2}{T_1 + T_2} \underset{k \in \mathcal{K}}{\sum} \| \mathbf{w}_k\|^2_2 \hspace{-1mm} + \hspace{-1mm} \frac{P}{(T_1 + T_2)v} \hspace{-1mm} \underset{\substack{m,n}}{\sum} \hspace{-1mm} \widetilde{\mathbf{d}}_{m,n}^T  \mathbf{z}_{m,n} \notag\\
\subto\hspace*{1mm} & \widetilde{\mbox{C1}}\mbox{:} \hspace{1mm} \widetilde{\gamma}_k \geq \Gamma_k, \hspace{1mm} \forall k \in \mathcal{K}, \notag \\
& \mbox{C2:}\hspace*{1mm} \left\| \bm{\alpha}_m \right\|^2_2 \leq 1, \hspace{1mm} \forall m, \notag \\
& \widetilde{\mbox{C3}} \mbox{:} \hspace*{1mm} \widetilde{\mathbf{x}}_m^T \mathbf{z}_{m,n} \hspace{-1mm} - \hspace{-1mm} \widetilde{\mathbf{x}}_m^T \mathbf{z}_{m,n-1} \hspace{-1mm} \geq \hspace{-1mm} \Delta_{\mathrm{min}}, \forall 2 \hspace{-1mm} \leq \hspace{-1mm} n \hspace{-1mm} \leq N, \forall m, \notag \\
& \widetilde{\mbox{C4}} \mbox{:} \hspace*{1mm} 0 \leq \widetilde{\mathbf{x}}_m^T \mathbf{z}_{m,n} \leq D, \hspace*{1mm} \forall n, \hspace{1mm} \forall m, \notag \\
& \widetilde{\mbox{C5}} \mbox{:} \hspace*{1mm} \widetilde{\mathbf{d}}_{m,n}^T \mathbf{z}_{m,n} \leq v T_1, \hspace*{1mm} \forall n, \hspace{1mm} \forall m, \notag \\
& \mbox{C11:}\hspace*{1mm} \mathbf{1}_{\widetilde{N}}^T \mathbf{z}_{m,n} = 1, \hspace*{1mm} \forall n, \hspace{1mm} \forall m, \notag \\
& \mbox{C12:}\hspace*{1mm} z_{m,n,\widetilde{n}} \in \{0, 1\}, \hspace*{1mm} \forall \widetilde{n}, \forall n, \hspace{1mm} \forall m.
\end{align}
\par
Compared to problem \eqref{continuous} for continuous antenna movement, problem \eqref{disc_h} has an additional binary constraint C12 for the antenna position selection variable, leading to a challenging MINLP problem.
\par
\subsection{Solution for Discrete Antenna Movement}
In this subsection, we develop a BCD-based algorithm to solve problem \eqref{discrete}.
The optimization variables are divided into two blocks, and the associated sub-problems are handled by capitalizing on bilinear transformation, penalty method, and SCA. A corresponding analysis of convergence and complexity is also provided.
\par
\begin{remark}
For continuous antenna movement, the channel coefficients depend on optimization variable $\mathbf{X}$ in problem \eqref{continuous} in a complicated manner, motivating the use of the variable splitting method, followed by using ADMM to handle the resulting equality constraints. However, problem \eqref{disc_h} for discrete antenna movement is formulated as an antenna position selection problem. The channel coefficients are a linear combination of the binary antenna position selection variables and the known channel coefficients of all possible antenna positions, making BCD suitable for handling the problem.
\end{remark}
\par
According to the principle of BCD, the variables of problem \eqref{discrete} are divided into two blocks, i.e., $\{ \mathbf{w}_k, \mathbf{Z}\}$ and $\{ \bm{\alpha}_m \}$. The sub-problems associated with the two blocks are solved in an alternating manner \cite{xu2025resolution}.
\par
\subsubsection{Update of $\{ \mathbf{w}_k, \mathbf{Z}\}$}
The sub-problem associated with block $\{ \mathbf{w}_k, \mathbf{Z}\}$ is given by
\begin{align} \label{bcd_1}
\hspace*{-3mm}\underset{\substack{ \mathbf{w}_k, \mathbf{Z} }}{\mino} \hspace*{3mm} & f_5  \notag\\
\subto\hspace*{3mm} & \widetilde{\mbox{C1}}, \widetilde{\mbox{C3}}\mbox{-}\widetilde{\mbox{C5}}, \mbox{C11}, \mbox{C12},
\end{align}
which is non-convex due to constraint $\widetilde{\mbox{C1}}$, the coupling between $\mathbf{w}_k$ and $\mathbf{Z}$, and binary constraint $\mbox{C12}$. In the following, we first transform $\widetilde{\mbox{C1}}$ into a more tractable form.
\par
We note that, similar to the transformation for \eqref{block_W}, a phase shift applied to $\mathbf{w}_k$ does not impact constraint $\widetilde{\mbox{C1}}$ nor the objective function value of problem \eqref{bcd_1}. Hence, $\widetilde{\mbox{C1}}$ can be equivalently expressed as
\begin{align}
\widetilde{\mbox{C1a}}\mbox{:} & \hspace{1mm} \left\lVert 
\begin{array}{c}
\mathbf{W}^H \mathbf{A}^H \mathbf{Z}^H \widetilde{\mathbf{h}}_k   \\ \sigma_k 
\end{array}
\right\rVert \leq \sqrt{1+\frac{1}{\Gamma_k}} \Re\left\{\widetilde{\mathbf{h}}_k^H \mathbf{Z} \mathbf{A} \mathbf{w}_k\right\}, \hspace{1mm} \forall k, \\
\widetilde{\mbox{C1b}} \mbox{:} & \hspace{1mm} \Im\left\{ \widetilde{\mathbf{h}}_k^H \mathbf{Z} \mathbf{A} \mathbf{w}_k \right\} = 0, \hspace{1mm} \forall k.
\end{align}
Next, we exploit the following bilinear transformation lemma to handle the variable coupling between $\mathbf{W}$ and $\mathbf{Z}$. 
\begin{lemma}
\textit{(Bilinear Transformation \cite{6698281})} The matrix equality $\mathbf{B} = \mathbf{C} \mathbf{D}$, where $\mathbf{C} \in \mathbb{C}^{M \times N}$, $\mathbf{D} \in \mathbb{C}^{N \times P}$, holds if and only if there exist $\mathbf{E} \succeq \mathbf{0}$, $\mathbf{F} \succeq \mathbf{0}$ such that the following two matrix inequalities hold
\begin{align}
& \begin{bmatrix}
\mathbf{E} & \mathbf{B} & \mathbf{C} \\
\mathbf{B}^H & \mathbf{F} & \mathbf{D}^H \\
\mathbf{C}^H & \mathbf{D} & \mathbf{I}_{N}
\end{bmatrix} \succeq \mathbf{0}, \label{bilinear_1} \\
& \mathrm{Tr}\left( \mathbf{E} - \mathbf{C}\mathbf{C}^H \right) \leq 0. \label{bilinear_2}
\end{align}
\end{lemma}
\par
Lemma 2 decouples the two coupled optimization variables using one convex constraint \eqref{bilinear_1} and one constraint \eqref{bilinear_2}, which is in the form of a d.c. function. Later, we will show that \eqref{bilinear_2} is actually convex by exploiting the special structure of binary selection variables.
Specifically, we introduce auxiliary variables $\mathbf{Q} \in \mathbb{C}^{M N \widetilde{N} \times K}$ with the $k$-th column being $\mathbf{q}_k$, and the following constraint
\begin{align}
\mbox{C13:} \hspace{1mm} \mathbf{Q} = \mathbf{ZAW}.
\end{align}
Based on Lemma 2, constraint C13 is equivalently transformed into the following two constraints
\begin{align}
& \mbox{C13a} \mbox{:} \hspace{1mm} 
\begin{bmatrix}
\mathbf{P} & \mathbf{Q} & \mathbf{Z} \\
\mathbf{Q}^H & \mathbf{S} & \mathbf{W}^H\mathbf{A}^H \\
\mathbf{Z}^H & \mathbf{A}\mathbf{W} & \mathbf{I}_{MN}
\end{bmatrix} \succeq \mathbf{0}, \\
& \mbox{C13b} \mbox{:} \hspace{1mm} \mathrm{Tr}(\mathbf{P} - \mathbf{Z}\mathbf{Z}^H) \leq 0,
\end{align}
where $\mathbf{P} \in \mathbb{C}^{M N \widetilde{N}} \succeq \mathbf{0}$ and $\mathbf{S} \in \mathbb{C}^{K} \succeq \mathbf{0}$ are auxiliary variables.
Moreover, we have 
\begin{align}
\mathrm{Tr}\left(\mathbf{Z} \mathbf{Z}^H\right) & = \underset{m \in \mathcal{M}}{\sum} \mathrm{Tr}\left(\mathbf{Z}_m \mathbf{Z}_m^H\right)  \notag \\
& = \underset{m \in \mathcal{M}}{\sum} \underset{n \in \mathcal{N}}{\sum} \mathbf{z}_{m,n}^H \mathbf{z}_{m,n} \notag \\
& = MN,
\end{align}
where the last equality is obtained from constraints C11 and C12.
As a result, constraint $\mbox{C13b}$ can be equivalently expressed as
\begin{align}
\widetilde{\mbox{C13b}} \mbox{:} \hspace{1mm} \mathrm{Tr}(\mathbf{P}) \leq MN.
\end{align}
Accordingly, constraints $\widetilde{\mbox{C1a}}$ and $\widetilde{\mbox{C1b}}$ are rewritten as
\begin{align}
\mywidetilde{\mbox{C1a}}\mbox{:} & \hspace{1mm} \left\lVert 
\begin{array}{c}
\mathbf{Q}^H \widetilde{\mathbf{h}}_k \\ \sigma_k 
\end{array}
\right\rVert \leq \sqrt{1+\frac{1}{\Gamma_k}} \Re\left\{\widetilde{\mathbf{h}}_k^H \mathbf{q}_k\right\}, \hspace{1mm} \forall k \in \mathcal{K}, \\
\mywidetilde{\mbox{C1b}} \mbox{:} & \hspace{1mm} \Im\left\{ \widetilde{\mathbf{h}}_k^H \mathbf{q}_k \right\} = 0, \hspace{1mm} \forall k \in \mathcal{K}.
\end{align}
Optimization problem \eqref{bcd_1} is now transformed into
\begin{align} \label{bcd_1_1}
\hspace*{-3mm}\underset{\substack{ \mathbf{w}_k, \mathbf{Z}, \\ \mathbf{Q}, \mathbf{P}, \mathbf{S} }}{\mino} \hspace*{3mm} & f_5  \notag\\
\subto\hspace*{3mm} & \mywidetilde{\mbox{C1a}}, \mywidetilde{\mbox{C1b}}, \widetilde{\mbox{C3}}\mbox{-}\widetilde{\mbox{C5}}, \notag \\
& \mbox{C11}, \mbox{C12}, \mbox{C13a}, \widetilde{\mbox{C13b}},
\end{align}
where the remaining non-convexity originates from binary constraint C12. To tackle this obstacle, we first transform constraint C12 equivalently into the following two constraints
\begin{align}
& \mbox{C12a:} \hspace{1mm} 0 \leq z_{m,n,\widetilde{n}} \leq 1, \hspace*{1mm} \forall \widetilde{n}, \hspace{1mm} \forall n, \hspace{1mm} \forall m, \\
& \mbox{C12b:} \hspace{1mm} \underset{\substack{m,n,\widetilde{n}}}{\sum} \left(- z^2_{m,n,\widetilde{n}} + z_{m,n,\widetilde{n}}\right) \leq 0,
\end{align}
where constraint C12b is a non-convex d.c. function. To deal with this issue, we first penalize it in the objective function by applying the penalty method, resulting in the following problem
\begin{align} \label{bcd_penalty}
\hspace*{-3mm}\underset{\substack{ \mathbf{w}_k, \mathbf{Z}, \\ \mathbf{Q}, \mathbf{P}, \mathbf{S} }}{\mino} \hspace*{3mm} & \overline{f}_5 \overset{\triangle}{=} f_5 + \zeta \underset{\substack{m,n,\widetilde{n}}}{\sum} \left(- z^2_{m,n,\widetilde{n}} + z_{m,n,\widetilde{n}}\right)  \notag\\
\subto\hspace*{3mm} & \mywidetilde{\mbox{C1a}}, \mywidetilde{\mbox{C1b}}, \widetilde{\mbox{C3}}\mbox{-}\widetilde{\mbox{C5}}, \notag \\
& \mbox{C11}, \mbox{C12a}, \mbox{C13a}, \widetilde{\mbox{C13b}},
\end{align}
where $\zeta > 0$ is the penalty parameter to enforce that constraint C12b is satisfied after optimization. 
An upper-bound on $\overline{f}_5$ is established as follows 
\begin{align}
\overline{f}_5 & \leq f_5 + \zeta \underset{\substack{m,n,\widetilde{n}}}{\sum} \left( \left(-2 \left(z_{m,n,\widetilde{n}}^{(i-1)}\right) + 1 \right)z_{m,n,\widetilde{n}} + \left(z_{m,n,\widetilde{n}}^{(i-1)}\right)^2 \right) \notag \\
& \overset{\triangle}{=} \widetilde{f}_5,
\end{align}
where the value of $z_{m,n,\widetilde{n}}^{(i-1)}$ is obtained in the previous iteration.
Optimization problem \eqref{bcd_penalty} is now transformed into
\begin{align} \label{bcd_1_final}
\hspace*{-3mm}\underset{\substack{ \mathbf{w}_k, \mathbf{Z}, \\ \mathbf{Q}, \mathbf{P}, \mathbf{S} }}{\mino} \hspace*{3mm} & \widetilde{f}_5  \notag\\
\subto\hspace*{3mm} & \mywidetilde{\mbox{C1a}}, \mywidetilde{\mbox{C1b}}, \widetilde{\mbox{C3}}\mbox{-}\widetilde{\mbox{C5}}, \notag \\
& \mbox{C11}, \mbox{C12a}, \mbox{C13a}, \widetilde{\mbox{C13b}},
\end{align}
which is a convex optimization problem and can be solved efficiently.
\subsubsection{Update of $\left\{ \bm{\alpha}_m \right\}$}
The sub-problem associated with $\left\{ \bm{\alpha}_m \right\}$ is a feasibility checking problem which is given by
\begin{align} \label{bcd_2}
\underset{\substack{ }}{\mathrm{Find}} \hspace*{3mm} & \bm{\alpha}_m   \notag\\
\subto\hspace*{3mm} & \widetilde{\mbox{C1}}, \mbox{C2}.
\end{align}
Similar to the steps for addressing problem \eqref{alpha_feasibility_continuous}, we update $\bm{\alpha}_m$ by solving the following optimization problem
\begin{align} \label{bcd_alpha_max}
\underset{\substack{ \bm{\alpha}_m }}{\maxo} \hspace*{3mm} & \underset{k \in \mathcal{K}}{\sum} \frac{\left| \widetilde{\mathbf{h}}_k^H\mathbf{Z} \mathbf{A} \mathbf{w}_k \right|^2}{ \underset{k' \in \mathcal{K}\setminus \{k\}}{\sum}\left| \widetilde{\mathbf{h}}_k^H\mathbf{Z} \mathbf{A} \mathbf{w}_{k'} \right|^2 + \sigma_k^2} \notag\\
\subto\hspace*{3mm} & \mbox{C2:} \hspace{1mm}  \left\| \bm{\alpha}_m \right\|^2_2 \leq 1, \hspace{1mm} \forall m \in \mathcal{M}.
\end{align}
Then, we recast optimization problem \eqref{bcd_alpha_max} as follows
\begin{align} \label{bcd_alpha_frac}
\underset{\substack{ \bm{\alpha}_m, \widetilde{\bm{\psi}} }}{\maxo} \hspace*{3mm} & f_6 \notag\\
\subto\hspace*{3mm} & \mbox{C2}, \mbox{C14},
\end{align}
where $\widetilde{\bm{\psi}} \in \mathbb{R}^{K \times 1}$ is an auxiliary variable whose $k$-th element is $\widetilde{\psi}_k$, and objective function $f_6$ is given by
\begin{align}
f_6 \overset{\triangle}{=} \underset{k \in \mathcal{K}}{\sum} \left( 2 \widetilde{q}_k \widetilde{\psi}_k - \widetilde{q}_k^2 \left( \underset{k' \in \mathcal{K}\setminus \{k\}}{\sum}\left| \widetilde{\mathbf{h}}_k^H \mathbf{Z} \mathbf{A} \mathbf{w}_{k'} \right|^2 + \sigma_k^2 \right) \right).
\end{align}
Furthermore, constraint C14 in \eqref{bcd_alpha_frac} is given by
\begin{align}
& \mbox{C14} \mbox{:} \hspace{1mm} \mathrm{Tr}\left( \left( \widetilde{\mathbf{T}}_k \mathbf{A}^{(i-1)} \mathbf{W}_k + \widetilde{\mathbf{T}}_k^T \mathbf{A}^{(i-1)} \mathbf{W}_k^T \right)^T \mathbf{A} \right) \notag \\
& - \mathrm{Tr} \left( \widetilde{\mathbf{T}}_k \mathbf{A}^{(i-1)} \mathbf{W}_k \left(\mathbf{A}^{(i-1)}\right)^T \right) - \widetilde{\psi}_k^2 \geq 0, \hspace{1mm} \forall k,
\end{align}
where $\widetilde{\mathbf{T}}_k \overset{\triangle}{=} \mathbf{Z}^H  \widetilde{\mathbf{h}}_k \widetilde{\mathbf{h}}_k^H \mathbf{Z} $. Variable $\widetilde{q}_k$ is updated according to
\begin{align}
\widetilde{q}_k = \frac{\left| \widetilde{\mathbf{h}}_k^H \mathbf{Z} \mathbf{A} \mathbf{w}_k \right| }{\underset{k' \in \mathcal{K}\setminus \{k\}}{\sum}\left| \widetilde{\mathbf{h}}_k^H \mathbf{Z} \mathbf{A} \mathbf{w}_{k'} \right|^2 + \sigma_k^2}, \hspace{1mm} \forall k \in \mathcal{K}.
\end{align}
\par
\begin{algorithm}[t]
\caption{BCD-based Algorithm for Solving Problem \eqref{discrete} for Discrete Antenna Movement}
\begin{algorithmic}[1]
\small
\STATE Set iteration index $i=1$, penalty parameter $\zeta > 0$, error tolerance $\widetilde{\varsigma}$. Initialize $\alpha_{m,n}^{(1)} = \frac{1}{\sqrt{N}}$ and $z_{m,n,\widetilde{n}}^{(1)} = 0$
\REPEAT
\STATE Update $\mathbf{w}_k$, $\mathbf{Z}$ by solving problem \eqref{bcd_1_final}
\STATE Update $\bm{\alpha}_m$ by solving problem \eqref{bcd_alpha_frac}
\STATE Set $i \leftarrow i + 1$
\UNTIL $\frac{\left|\widetilde{f}_5^{(i)} - \widetilde{f}_5^{(i-1)}\right|}{\widetilde{f}_5^{(i-1)}} \leq \widetilde{\varsigma}$
\end{algorithmic}
\end{algorithm}
\par
The BCD-based algorithm for solving problem \eqref{discrete} for discrete antenna movement is summarized in $\textbf{Algorithm 2}$. The objective function value of problem \eqref{bcd_1_final} is an upper-bound on the objective function value of problem \eqref{bcd_penalty}. This upper-bound is gradually tightened by iteratively solving problem \eqref{bcd_1_final}. Recall that solving \eqref{bcd_alpha_frac} does not directly influence the objective function value of \eqref{discrete}. Hence,
by iteratively solving problems \eqref{bcd_1_final} and \eqref{bcd_alpha_frac}, the objective function value of problem \eqref{discrete} is monotonically non-increasing. Thus, according to \cite{razaviyayn2013unified}, the proposed algorithm is guaranteed to converge to a stationary point of problem \eqref{discrete} within polynomial time.
The computational complexity of Algorithm 2 is dominated by solving problem \eqref{bcd_1_final} whose complexity is given by $\mathcal{O}( M^3N^3\widetilde{N}^3 K^{3.5} + MN\widetilde{N} K^{4.5} )$.
\vspace{-3mm}
\section{Simulation Results}
In this subsection, we validate the effectiveness of the proposed design of PASS via numerical simulations. We consider a square-shaped service area with side length of $D=40$ m. In particular, the PASS consists of $M = 3$ waveguides deployed at a height of $h = 5$ m, where each waveguide comprises $N = 4$ PAs, and $K = 3$ users are served. The $y$-axis coordinates of the three waveguides are $\{-15\ \text{m}, 0\ \text{m}, 15\ \text{m}\}$. 
The pinching antennas are preinstalled with the $x$-coordinate given by $\{8\ \text{m}, 16\ \text{m}, 24\ \text{m}, 32\ \text{m}\}$ for each waveguide. The system operates at a frequency of $f_c = 28$ GHz. The minimum antenna spacing is set to half of the wavelength \cite{10912473}. The effective refractive index of the waveguide is given by $n_{\text{eff}} = 1.4$ \cite{10945421}. The durations for antenna movement and signal transmission are given by $T_1 = 0.1$ s and $T_2 = 0.9$ s, respectively. The default motion power and antenna movement speed are set to $P = 0.1$ W and $v=1$ m/s, respectively \cite{Faulhaber}. The noise power and the SINR requirement of user $k$ are set to $\sigma_k^2 = -80$ dBm and $\Gamma_k = 24$, respectively. The penalty parameters are initialized as $\rho_1 = 10^3$ and $\rho_2 = 10^5$. The scaling factor is set to $\epsilon = 1.25$.
\par
For the purpose of performance comparison, we consider the following three baseline schemes.
\begin{itemize}
\item
\textbf{Baseline scheme 1:} This baseline scheme corresponds to the traditional MIMO design with the same number of RF chains as the considered PASS. Specifically, a MIMO BS equipped with $M=3$ antennas is deployed at the origin. 
The transmit power is minimized by optimizing the transmit beamforming while guaranteeing the SINR requirement of the users. The corresponding optimization problem can be equivalently transformed into a convex optimization problem and can be solved efficiently and optimally. The details are omitted here due to space limitations.
\item
\textbf{Baseline scheme 2:} This baseline scheme follows the existing PASS design without antenna radiation power control. In particular, the power is equally distributed among all the PAs of each waveguide, i.e., $\alpha_{m,n} = \sqrt{\frac{1}{N}}$. The resulting optimization problem can be solved by employing the proposed ADMM framework.
\item 
\textbf{Baseline scheme 3:} This baseline scheme exploits the existing PASS design that focuses on minimizing the information transmit power, i.e., $\frac{T_2}{T_1 + T_2} \underset{k \in \mathcal{K}}{\sum} \| \mathbf{w}_k\|^2_2$. The actual system performance is evaluated by the average power consumption, including the power consumption associated with antenna movement.
\end{itemize}
\par
For simplicity, in the following description, we refer to the proposed algorithms for continuous and discrete antenna movement as ``continuous design" and ``discrete design", respectively.
\begin{figure}[t]
\centering
\includegraphics[width=2.8in]{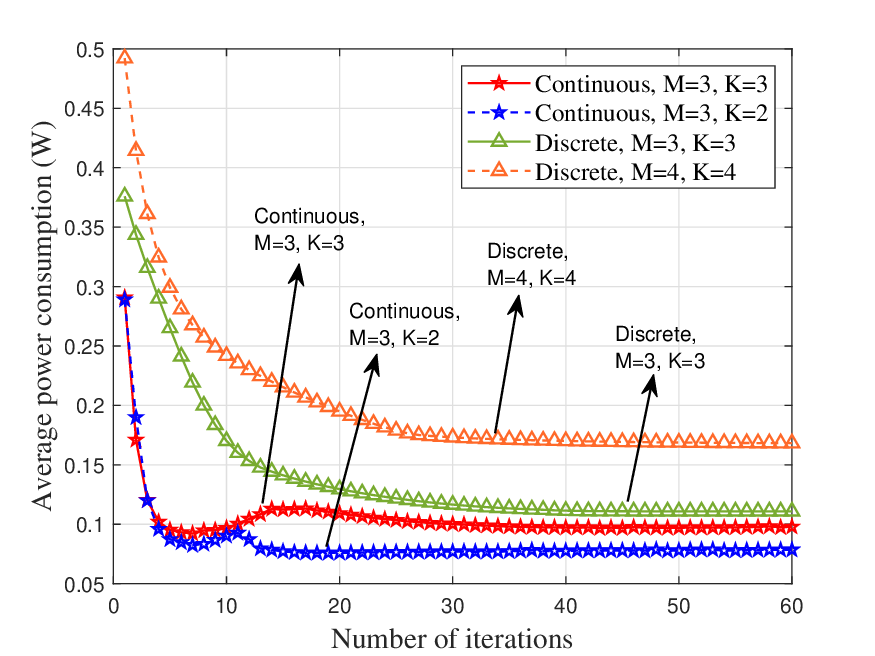}
\vspace*{-4mm}
\caption{Convergence behavior of the proposed algorithms for continuous and discrete antenna movement.}
\label{figure:Convergence}
\vspace*{0mm}
\end{figure}

\begin{figure}[t]
\centering
\includegraphics[width=2.8in]{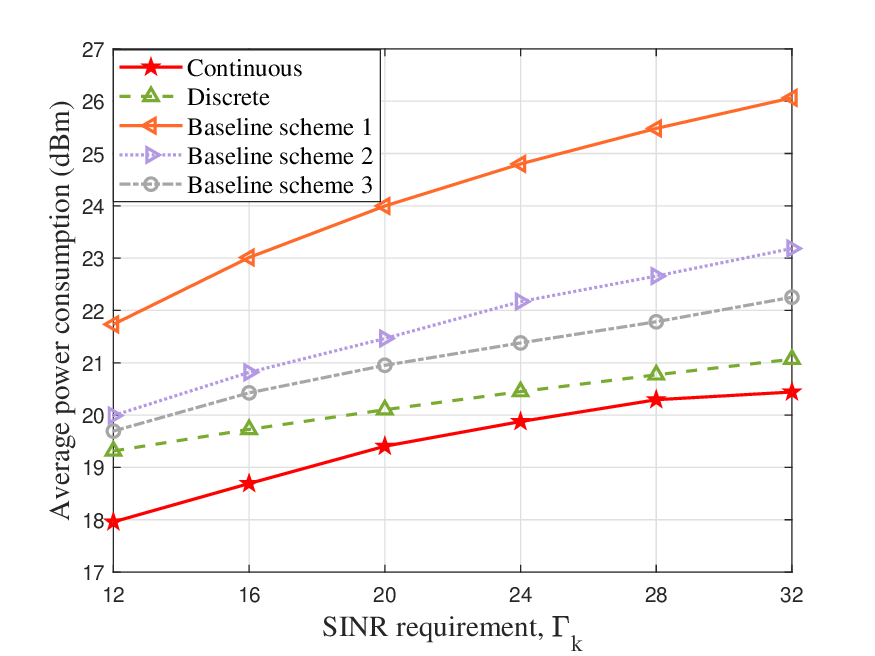}
\vspace*{-4mm}
\caption{Average power consumption (dBm) versus SINR requirement $\Gamma_k$.}
\label{figure:sinr}
\vspace*{0mm}
\end{figure}

\begin{figure}[t]
\centering
\includegraphics[width=2.8in]{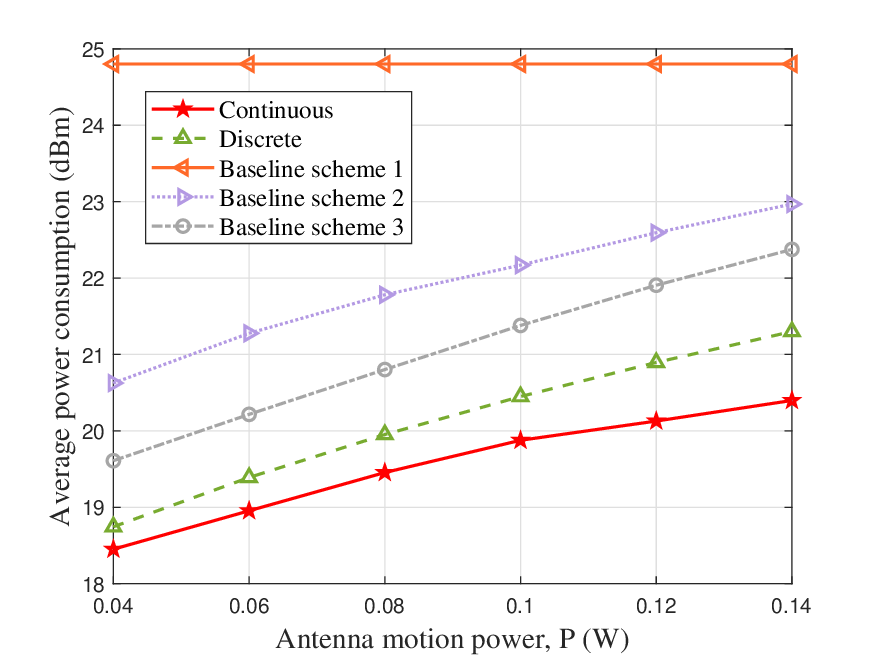}
\vspace*{-4mm}
\caption{Average power consumption (dBm) versus antenna motion power $P$ (W).}
\label{figure:motion_power}
\vspace*{-6mm}
\end{figure}

\begin{figure}[t]
\centering
\includegraphics[width=2.8in]{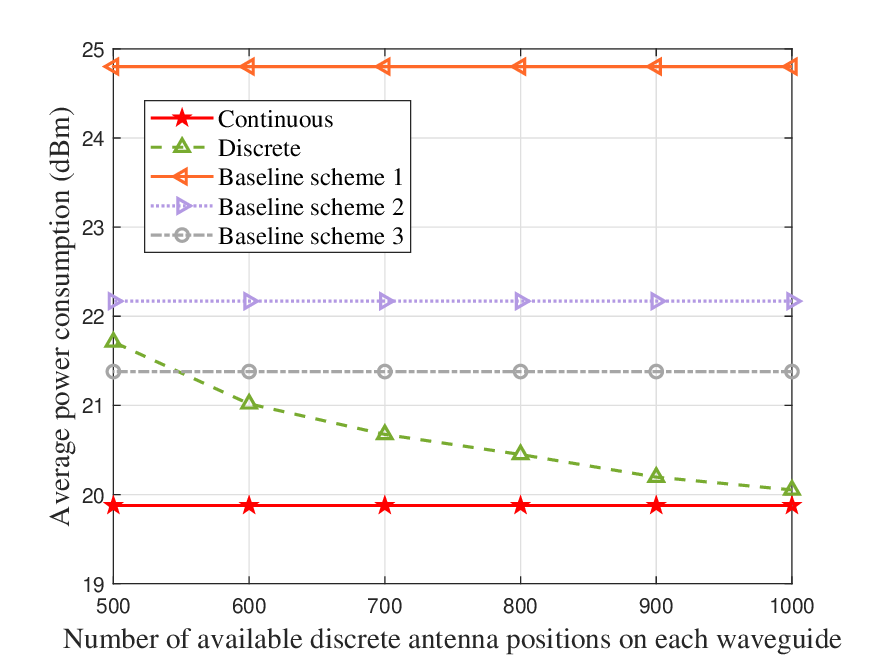}
\vspace*{-4mm}
\caption{Average power consumption (dBm) versus antenna spacing of discrete antenna positions.}
\label{figure:density}
\vspace*{0mm}
\end{figure}

\begin{figure}[t]
\centering
\includegraphics[width=2.8in]{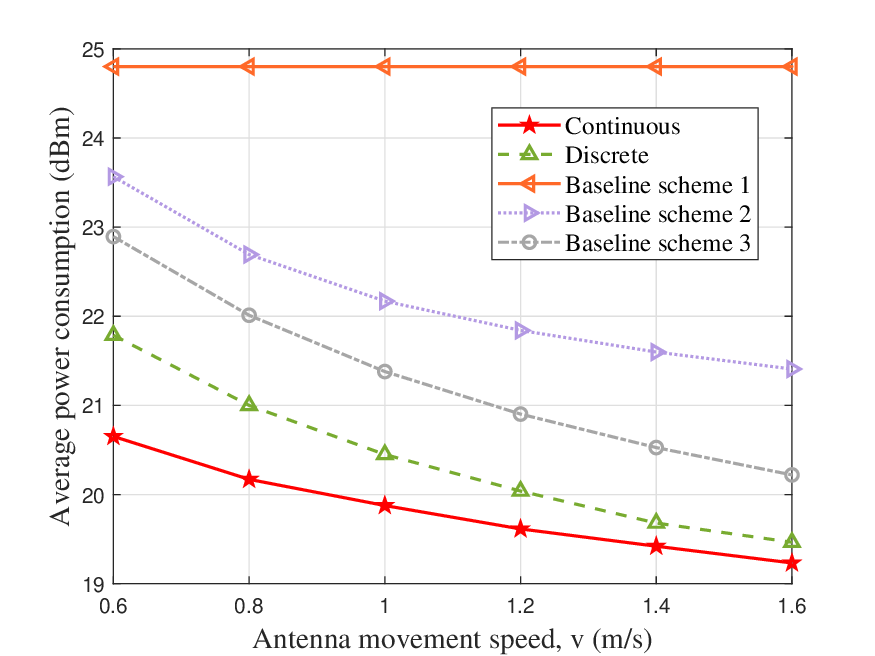}
\vspace*{-4mm}
\caption{Average power consumption (dBm) versus antenna movement speed $v$ (m/s).}
\label{figure:speed}
\vspace*{-5mm}
\end{figure}

\begin{figure}[t]
\centering
\includegraphics[width=2.8in]{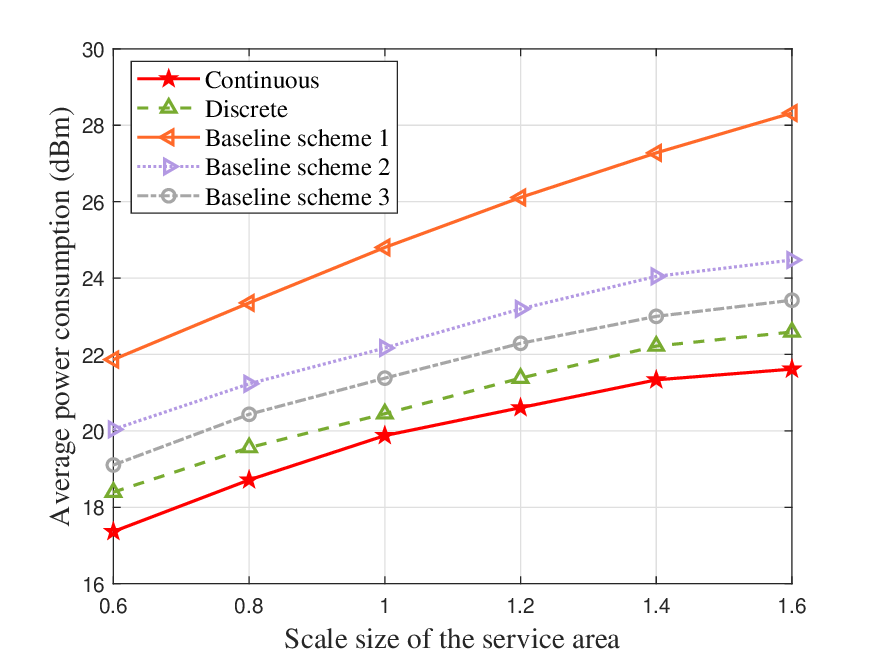}
\vspace*{-4mm}
\caption{Average power consumption (dBm) versus service area size.}
\label{figure:size}
\vspace*{-1mm}
\end{figure}

\vspace{-4mm}
\subsection{Algorithm Convergence}
Fig. \ref{figure:Convergence} presents the convergence behavior of the proposed algorithms. As can be observed, the average power consumption of Algorithm 1 for continuous antenna movement first decreases and then slightly increases, and finally decreases until convergence. This phenomenon is caused by the change of the penalty parameters. For the first several iterations, the penalty parameters are small, allowing large constraint violations and broadening the feasible set. Hence, the objective function value dramatically decreases, however, the solutions obtained during these initial iterations may not be feasible. As the penalty parameters increase, the constraint violations reduce gradually and the equality constraints are enforced, leading to a smaller feasible region and a larger objective function value. Once the equality constraints are satisfied, the objective function value decreases until convergence. As for Algorithm 2 for discrete antenna movement, the objective function value monotonically decreases until convergence. Furthermore, it can be observed that the algorithms require a few more iterations to converge as the number of antennas and users, i.e., $M$ and $K$, increases.
\vspace{-5mm}
\subsection{Impact of SINR Requirement}
We investigate the impact of the SINR requirements in Fig. \ref{figure:sinr}.
It can be observed from the figure that the average transmit power increases with the SINR requirement. Compared with baseline scheme 1, the proposed designs achieve significant performance improvement, which comes from the significantly reduced path loss in PASS. Compared with baseline scheme 2, the proposed design consumes less power because of the antenna radiation power optimization. This is because the flexible antenna radiation power facilitates more efficient energy utilization. In other words, the optimized antenna radiation power provides extra DoFs for system design and enlarges the feasible region of the optimization problem, leading to better system performance. The performance gap between the proposed scheme and baseline scheme 3 reveals that minimizing the information transmission power consumption alone cannot minimize the total power consumption in practice, thus verifying the necessity of also considering the power consumed for antenna movement.
\vspace{-4mm}
\subsection{Impact of Antenna Motion Power}
We study the impact of antenna motion power in Fig. \ref{figure:motion_power}. As can be observed, the average power consumption of PASS increases with the motion power consumption. The continuous design yields the lowest power consumption among all considered schemes.
As the antenna motion power increases from $0.04$ W to $0.14$ W, the continuous design also causes a smaller power consumption increment ($56.6\%$), compared to baseline scheme 3 ($89.1\%$). This is because, by taking the influence of the antenna motion power into account, the proposed scheme can strike a balance between the antenna motion power and information transmission power, thus reducing the average power consumption.
\vspace{-4mm}
\subsection{Impact of Density of Available Discrete Antenna Positions}
Fig. \ref{figure:density} illustrates the impact of the density of the available discrete antenna positions. We can observe that the average power consumption decreases with the number of available discrete antenna positions on each waveguide. This is because more available antenna positions support a finer-grained antenna movement control and hence improve performance. Furthermore, it can be observed that the discrete design approaches the continuous design when the number of available discrete antenna positions reaches $1000$, which verifies the optimality of the discrete design for the case of a large number of antennas. We note that 1000 antenna positions is a large number and results in a high computational complexity of the discrete design. On the other hand, although the discrete antenna movement case allows for easier hardware deployment, it suffers from a performance loss compared to the continuous design for a small number of antenna positions. As a result, the trade-off between system performance, computational complexity, and simplicity of hardware deployment of PASS with discrete antenna movement should be carefully considered for practical design.
\vspace{-4mm}
\subsection{Impact of Antenna Movement Speed}
The impact of the antenna movement speed is investigated in Fig. \ref{figure:speed}. As can be seen, the average power consumption decreases with the antenna movement speed. On the one hand, a larger speed expands the location range of the antennas, providing more feasible antenna positions for system design and enlarging the feasible region. On the other hand, a larger movement speed directly reduces the value of motion power consumption in the objective function. In addition, it can be seen from the figure that the performance gap between the discrete and continuous designs decreases with increasing antenna movement speed, as a higher speed reduces the impact of antenna motion power consumption in the objective function. However, a higher movement speed increases hardware costs. Hence, the antenna driver module should be carefully selected to balance hardware cost and system performance.

\vspace{-4mm}
\subsection{Benefit of PASS in Path Loss Mitigation}
In Fig. \ref{figure:size}, we investigate the performance advantage of PASS compared to conventional MIMO systems for varying service area sizes. In particular, we first define the scale size of the service area as $\frac{D'}{D}$, where $D'$ is the variable length of the square service area. As can be seen from the figure, the average power consumption increases with the service area size. Baseline scheme 1 consumes significantly more power as the service area increases, while the PASS is less sensitive to changes in the service area size. This is due to the fact that the PASS can alleviate the long-distance path loss via in-waveguide propagation. In addition, the PAs move to the preferred positions, facilitating short-range LoS communication. Fig. \ref{figure:size} demonstrates the significant path loss reduction and the efficient pinching beamforming achieved by PASS.


\vspace{-3mm}
\section{Conclusion}
This paper studied the design of PASS, where physical constraints on PA movement and the adjustable antenna radiation power were accounted for. The antenna positions, radiation power, and transmit beamforming were jointly optimized to minimize the average power consumption for both continuous and discrete antenna movement.
An ADMM-based framework was developed to tackle the unique challenges of PASS for continuous antenna movement by exploiting the unique ability of ADMM to address variable coupling and equality constraints. The problem for discrete antenna movement was recast as an MINLP and solved by the BCD method.
Simulation results showed that PASS offer significant benefits over conventional MIMO systems by greatly reducing path loss and alleviating user interference via subtle pinching beamforming. The system models proposed in this work, along with the derived optimization framework, represent an initial step toward fully harnessing the potential of PASS.



\vspace{-3mm}
\section*{Appendix}
For simplicity, we omit subscripts $k$, $m$, and $n$ in the derivation. We note that $\nabla_{\theta} v(\theta)$ is continuous and differentiable on $\mathbb{R}$. Then, according to the mean value theorem, for all $\theta_1$, $\theta_2 \in \mathbb{R}$, we have 
\begin{align}
\left| \nabla_{\theta} v(\theta_1) - \nabla_{\theta} v(\theta_2) \right| \leq \left|  \underset{\theta \in [\theta_1, \theta_2] }{\max} \nabla^2_{\theta} v(\theta) \right| | \theta_1 - \theta_2 |,
\end{align}
where the second-order gradient $\nabla^2_{\theta} v(\theta)$ is given by
\begin{align}
\nabla^2_{\theta} v(\theta) = 2 \beta \left|rt+\frac{\lambda}{\rho_2}\right| \cos \left( \theta - \angle\left( rt+ \frac{\lambda}{\rho_2} \right)\right),
\end{align}
which is upper bounded by $2 \beta \left|rt+\frac{\lambda}{\rho_2}\right|$. 
Hence, $\nabla_{\theta} v(\theta)$ is Lipschitz continuous with the Lipschitz constant given by $L = 2 \beta \left|rt+\frac{\lambda}{\rho_2}\right|$.

\vspace{-2mm}
\bibliographystyle{IEEEtran}
\bibliography{Reference_List}
\end{document}